\documentclass[twocolumn,tighten,trackchanges]{aastex631}
\usepackage{graphicx}		
\usepackage{url}
\usepackage{epstopdf}
\usepackage{color}
\usepackage{multirow}
\usepackage{yhmath}
\usepackage{ragged2e}
\usepackage{url}
\usepackage[utf8]{inputenc}
\usepackage{amsmath}

\def\lsim{\mathrel{\raise.3ex\hbox{$<$\kern-.75em\lower1ex\hbox{$\sim$}}}}
\def\gsim{\mathrel{\raise.3ex\hbox{$>$\kern-.75em\lower1ex\hbox{$\sim$}}}}
\def\gtwid{\mathrel{\raise.3ex\hbox{$>$\kern-.75em\lower1ex\hbox{$\sim$}}}}
\def\proptwid{\mathrel{\raise.3ex\hbox{$\propto$\kern-.75em\lower1ex\hbox{$\sim$}}}}
\DeclareUnicodeCharacter{2212}{-}

\def\apjl{ApJL}
\def\apjs{ApJS}
\def\ao{Appl.~Opt.}
\def\aap{A\&A}
\newcommand{\p}{\partial}

\newcommand{\suml}{\sum\limits}

\usepackage{hyperref}
\begin{document}

\title{Signatures of Black Hole Spin and Plasma Acceleration in Jet Polarimetry}
\shorttitle{Black Hole Spin from Jet Polarimetry}
\author{Z. Gelles}
\affil{Department of Physics, Princeton University, Princeton, NJ 08540, USA}
\author{A. Chael}
\affil{Princeton Gravity Initiative, Princeton University, Princeton, NJ 08540, USA}
\author{E. Quataert}
\affil{Department of Astrophysical Sciences, Princeton University, Peyton Hall, Princeton, NJ 08540, USA}

\begin{abstract}
We study the polarization of black hole jets on scales of $10-10^3\,GM/c^2$ and show that large spatial swings in the polarization occur at three characteristic distances from the black hole: the radius where the counter-jet dims, the radius where the magnetic field becomes azimuthally dominated (the light cylinder), and the radius where the plasma reaches its terminal Lorentz factor. To demonstrate the existence of these swings, we derive a correspondence between axisymmetric magnetohydrodynamic outflows and their force-free limits, which allows us to analytically compute the plasma kinematics and magnetic field structure of collimated, general relativistic jets. We then use this method to ray trace polarized images of black hole jets with a wide range of physical parameters, focusing on roughly face-on jets like that of M87.  We show that the location of the polarization swings is strongly tied to the location of the light cylinder and thus to the black hole's spin, illustrating a new method of measuring spin from polarized images of the jet. This signature of black hole spin should be observable by future interferometric arrays like the (Next Generation) Event Horizon Telescope, which will be able to resolve the polarized emission of the jet down to the near-horizon region at high dynamic range. 
\end{abstract}


\section{Introduction}
Black holes are believed to generate powerful outflows through a confluence of general relativistic and electrodynamic effects \citep{blandford_electromagnetic_1977,blandford_relativistic_1979,blandford_hydromagnetic_1982}. While the exact mechanism by which these jets are launched is still up for debate, the black hole's spin likely plays an important role, as magnetic fields threading the event horizon can convert the rotational energy of the hole into Poynting energy of the jet (BZ; \citealp{blandford_electromagnetic_1977}). The magnetic field structure of a black hole's jet is thus intimately connected to its near-horizon spacetime geometry. In this paper, we show that this connection leads to a strong signature of black hole spin that is observable via synchrotron polarization observations. 

The black hole jet in the elliptical galaxy Messier 87 (M87) is of particular importance, as it is one of the most visible radio sources in the local universe. Following the jet's discovery in 1918 \citep{curtis1918descriptions}, radio interferometry has proven to be an effective tool in mapping its shape and kinematics (e.g., \citealp{macdonald1968observations,junor1999formation,ly_high-frequency_2007,asada_structure_2012,hada_innermost_2013,hada_high-sensitivity_2016,park2019kinematics,kino_implications_2022}). In addition to the total intensity measurements, polarized Very-Long-Baseline-Interferometry (VLBI) observations of the M87 jet at 43 GHz and 86 GHz have also played an integral role in understanding its magnetic field structure \citep{hada_high-sensitivity_2016,walker_structure_2018,kravchenko2020linear}.

While the Event Horizon Telescope (EHT) has successfully resolved polarized emission all the way down to the near-horizon region at 230 GHz \citep{eht2021first}, this analysis cannot extend past a radius of $\sim 10GM/c^2$ due to the finite dynamic range of current observations. Conversely, observations of the large-scale jet at 86 GHz have been able to resolve polarized emission at a resolution of just $\sim 20GM/c^2$ \citep{hada_high-sensitivity_2016}, and observations at lower frequencies cannot probe the jet core due to increasing optical depth  \citep{lee_interferometric_2016}.

In the next decade, interferometric upgrades will make it possible to resolve the polarized emission from the jet at optically thin frequencies and at horizon-scale resolutions. In particular, joint efforts between the Global Millimetre VLBI Array (GMVA) and the Atacama Large Millimeter Array (ALMA) have begun to focus observations on the M87 jet at 86 GHz \citep{lu2023ring}, while the Next Generation Event Horizon Telescope (ngEHT) will perform similar analyses at 230 and 345 GHz \citep{chael2023multifrequency}. The addition of baselines extending to space, such as the Black Hole Explorer (BHEX), will even further increase the image resolution and allow for observations of more distant AGN \citep{johnson_black_2024}. 

With these future prospects in mind, we investigate the features of jet emission that will soon become visible for the first time. Our goals are two-fold: we aim to (1) develop a simplified model of relativistic outflows that can be computed analytically, and then (2) model polarized images of black hole jets, specifically demonstrating how they can be used to measure spin.

In the first half of the paper, we review the theoretical foundations of jet/wind launching and develop extensions of known solutions. Outflows have conventionally been discussed in the context of either Magnetohydrodynamics (MHD) or Force Free Electrodynamics (FFE). The former solves for the dynamics of plasma and the electromagnetic fields in tandem, while the latter treats the plasma as a negligible contribution to the system. Historically, theoretical developments in our understanding of jet launching have occurred separately between these two paradigms. 

In this paper, we will not restrict ourselves to the framework of MHD or FFE alone. Instead, we will establish a correspondence between the two, deriving force-free results as an exact limit of MHD and demonstrating the precise conditions under which the force-free condition holds. In particular, we will show that the force-free limit of cold, axisymmetric MHD predicts a unique plasma boost parallel to magnetic fieldlines, and that MHD Lorentz factors can be approximated by their force-free counterparts with a smooth cap (two novel results, to the best of our knowledge). 

In the second half of this paper, we then use these outflow solutions to compute polarized images of low-inclination black hole jets, bridging the near-horizon region (where the force-free approximation holds) to the far-field region (where full MHD is needed). Through analysis of these images, we show how three physical ingredients determine the variation of the polarization with distance from the black hole: (1) the relative importance of counter-jet vs. forward-jet inside the light cylinder, (2) the role of relativistic aberration as the magnetic field winds up outside the light cylinder and the plasma accelerates radially outwards, and (3)  the relative importance of plasma inertia vs. Poynting energy as the outflowing plasma reaches its terminal Lorentz factor. These three key components of spatially resolved jet polarimetry are exemplified in Figure~\ref{fig:annotatefig} of \S\ref{sec:discussion} and are explained in detail throughout the course of the paper.   

As this summary suggests, the light cylinder --- which corresponds to the point where the magnetic field transitions from poloidally dominated to azimuthally dominated --- plays a particularly important role in determining how the synchrotron polarization varies with distance from the black hole. We show that by measuring the polarization swing associated with the light cylinder in resolved jet polarimetry, one can measure the spin of the black hole. We expect this signature to be particularly effective because it is insensitive to a wide range of astrophysical complications. 

The rest of this paper is organized as follows. In \S\ref{sec:GRMHD}, we discuss the theoretical formalism of GRMHD outflows. In \S\ref{sec:ffsec}, we construct an accurate, analytic approximation for GRMHD outflows using the much simpler force-free formalism. In \S\ref{sec:imagesec}, we outline our model for computing polarized images of the jet. Readers less interested in the details of our formalism can proceed directly to \S\ref{sec:imageanalysis}, in which we present results for jet polarimetry as a function of distance from the black hole. These results are drawn together in \S\ref{sec:spinsec}, which details our proposed strategy for measuring spin. Finally, we conclude with a discussion of future model improvements in \S\ref{sec:discussion}.
\section{Stationary, Axisymmetric, Cold GRMHD}
\label{sec:GRMHD}
Stationary, axisymmetric MHD has served as an invaluable framework in analytically modelling astrophysical outflows. The flat-space MHD formalism was crucial in early attempts to model the solar wind \citep{weber_angular_1967}, pulsar magnetospheres \citep{mestel1979axisymmetric}, and winds launched from accretion disks \citep{blandford_hydromagnetic_1982}, after which the general relativistic extension was first presented in \cite{phinney_theory_1984,camenzind_hydromagnetic_1986,camenzind_hydromagnetic_1987}. In this section, we review the results of the general relativistic formalism, largely adopting the notation of \cite{phinney_theory_1984}. We then derive new critical conditions for jets with a range of magnetic field geometries.

\subsection{Setup}
\label{sec:setup}
The Kerr line element in Boyer-Lindquist coordinates $(t,r,\theta,\phi)$ is given by \citep{Kerr1963,chandrasekhar1998mathematical}
\begin{equation}
ds^2=-\frac{\Delta\Sigma}{\Xi}dt^2 +\frac{\Sigma}{\Delta} d r^2+\Sigma d\theta^2+\frac{\Xi\sin^2{\theta}}{\Sigma}\left(d \phi- \omega dt\right)^2, \\
\end{equation}
where
\begin{gather}
	\Delta=r^2-2Mr+a^2, \quad \Sigma=r^2+a^2\cos^2{\theta}, \nonumber \\
        \omega = \frac{2 a M r}{\Xi}, \quad \Xi = (r^2 + a^2)^2 - \Delta a^2 \sin^2\theta
\end{gather}
and $a=J/M^2$ is the dimensionless angular momentum of the black hole. Here and throughout the rest of this paper, we will work in natural units with $G=c=1$.  

In stationary, axisymmetric GRMHD, one begins by assuming the existence of a perfect fluid travelling with four-velocity $u^\mu$ and rest-frame energy density $\rho$, as well as electromagnetic fields that permeate the spacetime. We will take the ``cold" (zero temperature) limit, meaning that the pressure is zero and $\rho$ is proportional to the rest-frame number density of the fluid. 

The Faraday tensor $F_{\mu\nu}=\p_\mu A_\nu-\p_\nu A_\mu$ is built from the vector potential $A_\mu$. There are numerous conventions for defining the electromagnetic fields from $F$, which are all compared in \S\ref{sec:conventions}. In this paper, we will follow \cite{noble_primitive_2006} and define the fields as those viewed by the ``normal observer," which is the observer whose four-velocity is orthogonal to hypersurfaces of constant time:\begin{align}
\label{eq:normaleq}
    n_\mu\equiv(-\alpha,0,0,0),
\end{align}
with $\alpha\equiv (-g^{tt})^{-1/2}$ the lapse. In the context of the Kerr metric, the normal observer is often called the Zero-Angular-Momentum Observer (ZAMO). The electric and magnetic fields viewed by this observer are therefore
\begin{align}
\label{eq:emdef}
    \mathcal{E}^\mu\equiv -n_\nu F^{\nu\mu},\quad \mathcal{B}^\mu\equiv n_\nu (\star F)^{\nu\mu},
\end{align}  
where $(\star F)$ is the Faraday dual:\begin{align}
    (\star F)_{\mu\nu}&=\frac{1}{2}\epsilon_{\mu\nu\rho\sigma}F^{\rho\sigma}
\end{align}
with $\epsilon$ the Levi-Civita tensor\footnote{The Levi-Civita tensor is defined as\begin{align*}
    \epsilon_{\mu\nu\rho\sigma}&\equiv \sqrt{-g}[\mu\nu\rho\sigma],\quad
   \epsilon^{\mu\nu\rho\sigma}\equiv -\frac{[\mu\nu\rho\sigma]}{\sqrt{-g}},
\end{align*}
where $g$ is the metric determinant and $[\mu\nu\rho\sigma]$ is the totally antisymmetric symbol \citep{MTW_1973}.}.  

In axisymmetry, the $\phi$ component of the vector potential uniquely controls the poloidal fields, so it gets its own symbol $\psi \equiv A_\phi$ and is referred to as the ``stream function." Then the axisymmetric magnetic field can be written directly from the stream function and Faraday tensor as\begin{align}
    \mathcal{B}^t=0,\,\, \mathcal{B}^r=\frac{\alpha \psi_{,\theta}}{\sqrt{-g}},\,\, \mathcal{B}^\theta=-\frac{\alpha\psi_{,r}}{\sqrt{-g}},\,\, \mathcal{B}^\phi=\frac{\alpha F_{r\theta}}{\sqrt{-g}},
\end{align}
where $\sqrt{-g}=\Sigma\sin\theta$ is the metric determinant, and commas denote partial differentiation. Furthermore, it will be useful for us to define an additional quantity,\begin{align}
\label{eq:bbareq}
    \overline{B}\equiv (\star F)_{t\phi}=\alpha^{-1}\Delta\sin^2\theta \mathcal{B}^\phi,
\end{align}
which typically remains finite across the horizon even if $\mathcal{B}^\phi$ explodes. In flat space, $\overline{B}=r\sin\theta\sqrt{\mathcal{B}^\phi \mathcal{B}_\phi}$, so Amp\`ere's law implies that $\overline{B}$ will be proportional to the current sourcing the azimuthal field (this will become relevant in \S\ref{sec:bfieldff}).
We note that $\overline{B}$ is referred to as $``B_T"$ in \cite{phinney_theory_1984}. 

From here, all dynamical variables can be neatly combined into the stress tensor\begin{align}
    T^{\mu\nu}_{\rm MHD}&=T^{\mu\nu}_{\rm EM}+T_{\rm Fluid}^{\mu\nu},
\end{align}
where in Heavisize-Lorentz units,
\begin{align}
       T^{\mu\nu}_{\rm EM}&=F^{\mu\rho}F^\nu_{\,\,\,\rho}-\frac{1}{4}g^{\mu\nu}F^2,\quad T^{\mu\nu}_{\rm Fluid}=\rho u^\mu u^\nu.
\end{align}
In the definition of $T^{\mu\nu}_{\rm MHD}$, we have neglected the contribution from pressure in the cold limit. Finally, in MHD, one demands that the electric field vanish inside the fluid frame, which is equivalent to the requirement that the fluid have infinite conductivity.  

The system is then described by five sets of covariant equations \citep{komissarov_godunov-type_1999,gammie_harm_2003}:\begin{align*}
    &1.\,\text{Energy-Momentum Conservation}:\nabla_\mu T^{\mu\nu}_{\rm MHD}=0.\\
    &2. \,\text{Maxwell}:\nabla_\mu(\star F)^{\mu\nu}=0.\\
    &3.\, \text{Continuity}:\nabla_\mu(\rho u^\mu)=0.
    \\
    &4.\, \text{MHD Condition}: u_\mu F^{\mu\nu} =0.
    \\
    &5.\, \text{Normalization}:u^\mu u_\mu=-1.
\end{align*}
In axisymmetry, the plasma dynamics can be re-cast in terms of four conserved quantities: a mass flux $\eta$, a fieldline rotation rate $\Omega_F$, a specific energy $E$, and a specific angular momentum $L$. These quantities are conserved in the sense that they are constant along contours of fixed $\psi$, and they are explicitly given by\footnote{In Gaussian units, $\eta\to 4\pi\eta$.} (see Appendix~\ref{sec:appconserve} for derivation and additional comments)\begin{align}
\label{eq:etaeq}
    \eta&=\frac{\alpha\rho u^r}{\mathcal{B}^r}=\frac{\alpha\rho u^\theta}{\mathcal{B}^\theta}, &&\Omega_F=\frac{F_{tr}}{F_{r\phi }}=\frac{F_{t\theta}}{F_{\theta\phi}}
    \\
    \nonumber
    E&=-u_t-\frac{\Omega_F\overline{B}}{\eta}, &&\,\,\,L=u_\phi-\frac{\overline{B}}{\eta}.
\end{align}
These conserved quantities fix the dynamics of the fluid in the spacetime.

\subsection{Electromagnetic Fields}
While conserving the above quantities is a requirement for the MHD equations to be satisfied, they do not constrain the stream function $\psi$. Ultimately, Maxwell's equations and energy-momentum conservation can be condensed into a single Grad-Shafranov (GS) equation for the stream function $\psi$, which encapsulates the force balance between the electromagnetic stresses and the matter. But the full GS equation in curved space is very lengthy \citep{nitta_effects_1991} and admits few analytical solutions, so we will turn to approximations. 

We will fix the stream function\footnote{This expression describes the stream function in the northern hemisphere $(0\leq \theta\leq \pi/2)$. To ensure that the no-monopole constraint holds at the origin, we ``split" the field configuration in the southern hemisphere ($\cos\theta\to-\cos\theta$), as is standard practice.} (see, e.g., \citealp{tchekhovskoy_simulations_2008,Broderick_2009sil})\begin{align}\label{eq:streamfunc}
    \psi&= C r^p \left(1-\cos\theta\right),
\end{align}
where $C$ is a constant and $0\leq p<2$ is a collimation parameter. To ensure that $\psi$ has the correct dimensions, we can break down the constant as\begin{align}
\label{eq:constdefn}
    C&=\Omega_F^{p-2}\eta\sigma_M,
\end{align}
where following \cite{michel_relativistic_1969}, $\sigma_M$ is a dimensionless parameter that controls the strength of the magnetic field. In particular, when $\sigma_M$ is large and we enter the force-free regime (described in \S\ref{sec:ffsec}), Eq.~\ref{eq:streamfunc} becomes an approximate solution to the flat-space GS equation \citep{tchekhovskoy_simulations_2008}. 

\begin{figure}[h]
    \centering
    \includegraphics[width=.4\textwidth]{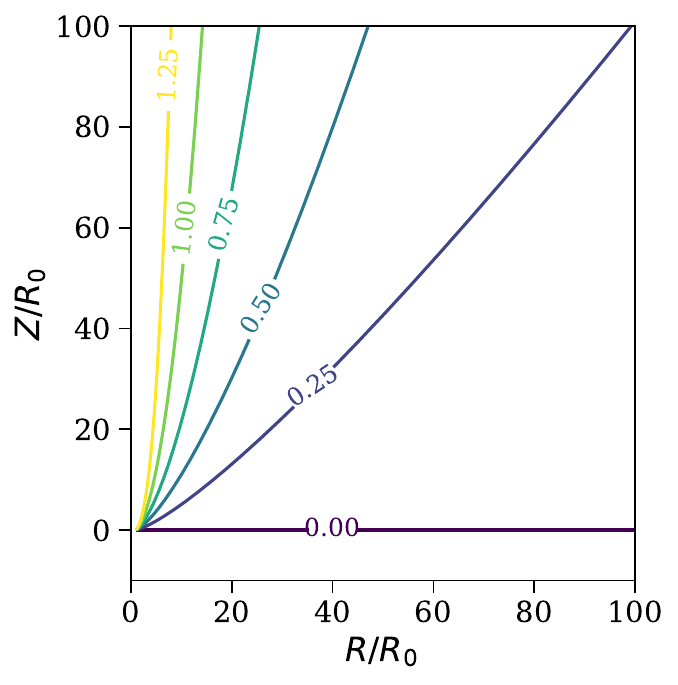}
    \caption{Shape of poloidal fieldlines plotted in terms of contours of constant $p$ (as labeled), which is a collimation parameter that controls the steepness of the fieldlines. Axes are cylindrical coordinates $R$ and $Z$, which are expressed in terms of the launch radius $R_{0}$.}
    \label{fig:fieldlineshape}
\end{figure}

From this choice of vector potential, the poloidal magnetic fields are given by\begin{align}
\label{eq:magb}
   \mathcal{B}^r&=\frac{\alpha\eta r^p\Omega_F^{p-2}\sigma_M}{\Sigma}
   \\
   \mathcal{B}^\theta&=-\frac{p\alpha\eta r^{p-1}\Omega_F^{p-2}\sigma_M(1-\cos\theta)}{\Sigma\sin\theta}.
\end{align}
To visualize the shape of the fieldlines, it is useful to switch to dimensionless cylindrical coordinates\footnote{We convert to dimensionless coordinates using the length scale $\Omega_F^{-1}$, rather than $M$, so that the coordinates do not degenerate in the flat space limit $M\to 0$.}\begin{align}
    R\equiv r\Omega_F\sin\theta, \qquad Z\equiv r\Omega_F\cos\theta.
\end{align}
For $Z\gg 1$ (far from the black hole), one can show that along a contour of fixed $\psi$, the coordinates are related as (see Appendix~\ref{sec:critapp2})\begin{align}
\label{eq:scaleeq}
    Z\propto R^{\frac{2}{2-p}}.
\end{align}
Thus the vector potential in Eq.~\ref{eq:streamfunc} defines a collimated field geometry, with $p=0$ corresponding to monopole and $p=1$ corresponding to paraboloid. Contour plots of the fieldline shape for various values of $p$ are shown in Figure~\ref{fig:fieldlineshape}.  The collimation of the field lines determines how much of the relativistic motion of the plasma is directed towards the observer, and thus the effects of relativistic aberration on the emission (see \S \ref{sec:relabb}).

Black hole magnetospheres found in numerical simulations match this particular form of the stream function with $p\simeq 0.75$ \citep{mckinney_disk-jet_2007,tchekhovskoy_simulations_2008,nakamura_parabolic_2018}, although the precise value of $p$ is likely spin dependent and could be larger than $1$ in certain scenarios \citep{narayan_jets_2022}. VLBI observations reveal a similar phenomenon, with studies of M87 and NGC 315 finding data consistent with $p\simeq 0.75$ as well \citep{asada_structure_2012,hada_innermost_2013,nakamura_parabolic_2018,park_jet_2021,park_discovery_2024}. 

\subsection{Winds}
\label{sec:winds}
Here, we describe how to use the above formalism to solve for the dynamics of the fluid along fieldlines, and we discuss the conditions under which wind outflows can be launched. 

Once the background stream function and conserved quantities are fixed, one can impose the $r$ component of the MHD condition to solve for $\overline{B}$ as a function of $u^\mu$ and the conserved quantities. This in turn can be inverted from Eq.~\ref{eq:etaeq} so that $u_t$ and $u_\phi$ are explicit functions of $E$, $L$, $u^r$, and $u^\theta$. After imposing $u^\theta=\frac{\mathcal{B}^\theta u^r}{\mathcal{B}^r}$ (which follows from conservation of $\eta$), we have just one remaining component of the four-velocity to compute: $u^r$. Its solution comes from the normalization $u^\mu u_\mu=-1$. In fact, it turns out that this normalization constraint can be written in terms of the conserved quantities as a quartic polynomial in $u^r$,
which is known as the wind equation \citep{phinney_theory_1984,camenzind_hydromagnetic_1986,tomimatsu_relativistic_2003} and is derived in more detail in Appendix~\ref{sec:windeqn}.  

The roots of the wind equation, which are most easily calculated numerically, thus determine the dynamics of the plasma. This process of numerical root-finding to compute outflows in curved space has been utilized many times, originally being applied to monopolar accretion \citep{camenzind_hydromagnetic_1986,takahashi_magnetohydrodynamic_1990,gammie_efficiency_1999}, as well as more collimated fieldline geometries in the Kerr spacetime \citep{nitta_effects_1991}.

The outflows are launched when solutions to the wind equation accelerate $u^r$ from zero to positive values as $r$ increases. By the frozen-flux theorem (see \citealp{gralla_spacetime_2014} for a simple proof in curved space), these winds will be confined to constant poloidal fieldlines, though nonzero angular momentum will allow them to spiral in $\phi$. In order to generate such an outflow, several conditions must be met, which we present below. 

\subsubsection{Conditions for Wind Launching}

First, at the point where the wind is launched (often referred to as the ``footpoint"), the plasma must be co-rotating with the fieldlines (see Eq.~\ref{eq:launcheq2} for derivation):
\begin{align}
\label{eq:launcheq}
    u^\mu_{0}\propto (1,0,0,\Omega_F).
\end{align}
This is only possible when the four-vector $(1,0,0,\Omega_F)$ is timelike. The cylindrical radii where this four-vector becomes null are respectively referred to as the \emph{inner and outer light cylinders} $R_{L,\rm in}$ and $R_{L,\rm out}$ \citep{goldreich_pulsar_1969,phinney_theory_1984}, and are given by the condition:\begin{align}
\label{eq:lightcyl}
    (u_\mu^{0})^2|_{R=R_L}=-N_{\rm co}|_{R=R_L}=0,
\end{align}
where we have defined the normalization factor\begin{align}
\label{eq:ncofac}
N_{\rm co}\equiv -(g_{tt}+2g_{t\phi}\Omega_F+g_{\phi\phi}\Omega_F^2).
\end{align}
So for a wind to be generated, the launch point must lie between $R_{L,\rm in}$ and $R_{L,\rm out}$. 

One can show that the outer light cylinder scales as $R_{L,\rm out}\sim \Omega_F^{-1}$, which reflects the fact that the co-rotating frame becomes null when the fieldlines ``move" too quickly. The inner light cylinder, on the other hand, can be qualitatively understood from the reverse perspective: sufficiently close to the event horizon, the frame-dragging of the black hole sweeps timelike observers around faster than the fieldlines. In the limits $M\to 0$ or $a\to 0$, one has $R_{L,\rm in}\to 0$. 

Between the inner and outer light cylinders, one can show that the magnetic field will be predominantly poloidal ($r$ and $\theta$). At the outer light cylinder, however, the fieldlines begin to rotate faster than the plasma, causing the magnetic field to ``wind up" and become predominantly azimuthal \citep{spruit_magnetohydrodynamic_1996}. In this manner, the light cylinder serves as an important dynamical surface that separates region of strong poloidal fields from strong azimuthal fields. As a result, the light cylinder will have a strong imprint on the polarization signatures of relativistic jets, as we show in \S\ref{sec:imageanalysis}.

Second, the footpoint must be located outside of the \emph{stagnation surface}, which defines a boundary between inflows that are launched towards the black hole and winds that are launched away from the black hole. In Appendix~\ref{sec:stag}, we show (in agreement with \citealp{takahashi_magnetohydrodynamic_1990}) that the stagnation surface $R_{\rm stag}$ satisfies\begin{align}
\label{eq:stageq}
   \mathcal{B}^\mu\nabla_\mu N_{\rm co}|_{R=R_{\rm stag}}=0.
\end{align}
In other words, the stagnation surface occurs at the point where $N_{\rm co}$ is a local extremum along a fieldline. 

The stagnation surface corresponds to the point where gravitational and centrifugal forces balance. Due to frame dragging, the location of the stagnation surface will depend sensitively on spin too: sufficiently close to the event horizon, the proper angular speed of the co-rotating frame can \emph{decrease} with $r$, causing centrifugal forces to launch material inwards rather than outwards. This generalizes the non-relativistic picture of centrifugally launched winds described by, e.g., \cite{spruit_magnetohydrodynamic_1996}, and also builds on the 
discussions of rotating, relativistic disks in  \cite{abramowicz_relativistic_1978,penna_new_2013}.

Third, one must choose appropriate values of $E,L$ and $\sigma_M$ that ensure the wind equation has a real root. We discuss how to do this in the following section.

\subsection{Critical Point Analysis}
\label{sec:critsec}
To determine the values of $E$, $L$, and $\sigma_M$ that produce physically sensible winds, one typically imposes additional algebraic constraints that follow from regularity at the critical (singular) points \citep{goldreich_stellar_1970,phinney_theory_1984,camenzind_hydromagnetic_1986}. 
The critical points are locations in the flow where the poloidal velocity of the matter matches the phase velocity of the plasma waves (either magnetosonic or Alfv\'en). The regularity conditions at the critical points are discussed in detail in Appendix~\ref{sec:critapp0}, and we review the main results here.

The location of the fast point --- where the poloidal velocity of the flow matches the fast magnetosonic speed --- constrains the conserved quantity $E$. In this work, we will assume that the fast point is located at $R=\infty$, where the spacetime is flat and the regularity conditions simplify. If the fast point were located a finite radius instead, the critical conditions would be very complicated and require numerical root finding (see, e.g., \citealp{li_electromagnetically_1992, takahashi_trans-fast_1998,takahashi2002transmagnetosonic}). In \S\ref{app:finitefms}, we discuss the dynamical effects of a finite fast point, and we explain why such a change to the model would not alter the conclusions of our paper. 

In the case of the monopole, the critical conditions for an infinite fast point constitute a well-studied problem and are described by Michel's ``minimum-energy" solution for winds that reach spatial infinity \citep{michel_relativistic_1969, goldreich_stellar_1970}. The model is analogous to the \cite{weber_angular_1967} solar wind model and smoothly converts Poynting energy into kinetic energy. The critical conditions for such a monopolar solution are:\begin{align}
\label{eq:monocrit}
    E|_{p=0}=\gamma_\infty^3,\quad \sigma_M|_{p=0}=\frac{(\gamma_\infty^2-1)^{3/2}}{\sin^2\theta},
\end{align}
where $\gamma_\infty$ is the terminal Lorentz factor and $\theta$ is the polar angle of the fieldline.  

However, the conditions for an infinite fast point will be different in stream functions with $p>0$, as the fieldlines become asympotically vertical in that case. We derive the $p>0$ critical conditions under this assumption in Appendix~\ref{sec:critapp2}. The result is that\begin{align}
\label{eq:parcrit}
    E|_{p>0}=\gamma_\infty^{3},\quad \sigma_M|_{p>0}=\frac{\sin^p\theta_0(\gamma_\infty^2-1)^{3/2}}{2R_{0}^p(1-\cos\theta_{0})},
\end{align}
where $R_{0}$ is the dimensionless cylindrical radius from which the wind is launched. As far as we are aware, Eq.~\ref{eq:parcrit} has not been derived elsewhere in the literature, though several papers have demonstrated the ubiquity of $E\propto \gamma_\infty^3$ (see Appendix~\ref{app:finitefms}). 

The fact that Eqs.~\ref{eq:monocrit} and \ref{eq:parcrit} do not agree as $p\to 0$ arises precisely from the discontinuous change in the asymptotic collimation profile. In any case, we see that we can map the obscure quantities $E$ and $\sigma_M$ onto the physical observables $\gamma_\infty$ and $R_{0}$. 

We still need, however, to find a physical observable to map onto $L$. It turns out that $L$ can also be written in terms of the launch point $R_{0}$. Taking $u^r=0$ in the wind equation, we show in Appendix~\ref{sec:windeqn} that\begin{align}
\label{eq:constraint}
    L&=\Omega_F^{-1}(E-\sqrt{N_{\rm co}})|_{R=R_0}.
\end{align}
Thus $E,L,$ and $\sigma_M$ all have clean, physical interpretations in this infinite fast point model. 

In any case, one has the freedom to choose any value of $\gamma_\infty$ and $R_{0}$ to determine $E$ and $L$, so long as $\gamma_\infty>1$ and the relevant surfaces lie in the correct order:\begin{align}
    R_{L,\rm in}<R_{\rm stag}\leq R_{0}<R_{L,\rm out}.
\end{align}
For a wind to smoothly transition from inflow to outflow, the launch point must be located precisely at the stagnation surface, so $R_0=R_{\rm stag}$. This will be our default assumption going forward.

\section{Force-Free Approximations to GRMHD Winds}
\label{sec:ffsec}
In this section, we show how the algebraically simpler force-free  winds can be used to derive excellent approximations to the cold GRMHD winds analyzed in the previous section. In particular, we show that the force-free limit of GRMHD predicts \emph{unique} and \emph{analytic} values for the azimuthal field $\overline{B}$, the fieldline rotation $\Omega_F$, and the plasma velocity $u^\mu$. We then derive an explicit correspondence between GRMHD and FFE that demonstrates how the force-free results can be used to analytically approximate their MHD counterparts, so long as the outflow is moderately relativistic (terminal Lorentz factor greater than a few). 

\subsection{Force-Free Electrodynamics}
In force-free electrodynamics, the fluid is neglected completely, and all stresses are assumed to come from the electromagnetic fields: \begin{align}
\label{eq:stressFF}
    T^{\mu\nu}_{\rm MHD}\Rightarrow T^{\mu\nu}_{\rm EM}.
\end{align}
Since $\nabla^\mu T_{\mu\nu}^{\rm EM}=J^{\mu}F_{\mu\nu}$ always, then Eq.~\ref{eq:stressFF} implies that \begin{align}
  J^\mu F_{\mu\nu}=0
\end{align}
in FFE,
which is known as the force-free condition; this determines the currents in terms of the fields. The equations of motion for the plasma in this regime can be obtained by taking the limit of MHD with $\sigma_M\to\infty$ and $\eta\sim\sigma_M^{-1}\to 0$: this ensures that the stresses are purely electromagnetic while keeping the stream function (which is proportional to $\eta\sigma_M$ via Eq.~\ref{eq:constdefn}) finite.

Indeed, the stream functions of the form Eq.~\ref{eq:streamfunc} become approximate solutions to the GS equation in the force-free limit for all $0\leq p<2$ \citep{tchekhovskoy_simulations_2008}. The monopole stream function $\psi\propto 1-\cos\theta$ also becomes an \emph{exact} solution in Schwarzschild \citep{michel_rotating_1973},
with MHD corrections going like $\sigma_M^{-2/3}\sim \gamma_\infty^{-2}$ \citep{beskin_mhd_1998}.

The force-free limit has been particularly useful in describing pulsar magnetospheres \citep{mestel1973force,contopoulos_axisymmetric_1999,uzdensky_axisymmetric_2003}, as well as near-horizon Kerr black hole magnetospheres \citep{blandford_electromagnetic_1977,thorne_electrodynamics_1982,okamoto_force-free_2009}. In these strongly gravitating, highly magnetized environments, the force-free condition is expected to approximately hold everywhere but the equatorial plane, where a thin current sheet sources the electromagnetic fields \citep{uzdensky_forcefree_2005,mckinney_discjet_2007}. Indeed, numerical simulations of FFE have proven to be an important tool in a wide range of astrophysical environments, as the FFE formalism both reduces computational expense and avoids the numerical difficulties of GRMHD in Poynting dominated regions \citep{mckinney_general_2006-1,tchekhovskoy_simulations_2008,lehner_intense_2012,chael2024hybrid}.

Force-free solutions are particularly concise and predict unique values for all dynamical variables, as we demonstrate in the subsequent subsections.  

\subsection{Azimuthal Magnetic Field}
\label{sec:bfieldff}
First, we will show that in the force-free limit, the azimuthal magnetic field $\overline{B}$ becomes a constant that is independent of the field collimation profile for all $p>0$. To see this, we start by solving Eq.~\ref{eq:etaeq} for $\overline{B}$:\begin{align}
    \overline{B}&=-\frac{\eta E}{\Omega_F}-\frac{u_t\eta}{\Omega_F}.
\end{align}
In the force-free limit, $\eta\to 0$ while $\eta E\sim\eta \sigma_M$ remains finite, meaning we can disregard the second term and hence\begin{align}
\label{eq:bbarconstant}
    \overline{B}|_{\rm FF}&=-\frac{\eta E}{\Omega_F}={\rm const},
\end{align}
in agreement with Eq.~185 of \cite{camenzind_hydromagnetic_1987}.
The nature of this constant depends on the precise value of the stream function $\psi$. 

Thus we see that another conserved quantity, $\overline{B}$, emerges in the force-free limit. It is conventional to normalize this conserved quantity by a factor of\footnote{In Gaussian units, $I=\overline{B}/2$.} $2\pi$, which gives a ``poloidal current" 
that sources the azimuthal field via Amp\`ere's law \citep{contopoulos_axisymmetric_1999,gruzinov_power_2005,gralla_spacetime_2014}:\begin{align}
    I&\equiv 2\pi (\star F)_{t\phi}=2\pi \overline{B}.
\end{align}
Let us now compute this poloidal current for the magnetic field geometries of interest. To do so, we will impose the MHD critical conditions (which still hold in the force-free limit) to compute $E$, and then substitute into Eq.~\ref{eq:bbareq}. For $p=0$, we use the critical condition in Eq.~\ref{eq:monocrit} to see that when a monopolar wind is launched along a fieldline of polar angle $\theta$, one gets\begin{align}
E|_{p=0}\approx\sigma_M\sin^2\theta=\frac{\psi\sin^2\theta\Omega_F^2}{\eta(1-\cos\theta)}=\frac{\psi\Omega_F^2(1+\cos\theta)}{\eta},
\end{align}
and hence a poloidal current of\begin{align}
\label{eq:monoI}
I=2\pi \overline{B}=-2\pi\frac{\eta E}{\Omega_F}=-2\pi\psi\Omega_F(1+\cos\theta),
\end{align}
in agreement with the \cite{michel_rotating_1973} solution.

For $p>0$, on the other hand, we can plug in from Eq.~\ref{eq:parcrit}:\begin{align}
\label{eq:powerI}
    E|_{p>0}\approx 2\left(\frac{R_{0}}{\sin\theta_0}\right)^p(1-\cos\theta_{0})\sigma_M=\frac{2\psi\Omega_F^2}{\eta},
\end{align}
which gives a poloidal current of\begin{align}
\label{eq:ffpar}
    I|_{p>0}&=-4\pi\psi\Omega_F.
\end{align}
For $p=1$, this matches the paraboloidal poloidal current predicted by \cite{blandford_accretion_1976} at large radii. Here, we have shown from MHD that it holds for all $0<p<2$. This principle agrees with Eq. 35 of \cite{nathanail_black_2014}, who showed that the same result follows from the force-free GS equation. 

\begin{figure}[h]
    \centering
    \includegraphics[width=.48\textwidth]{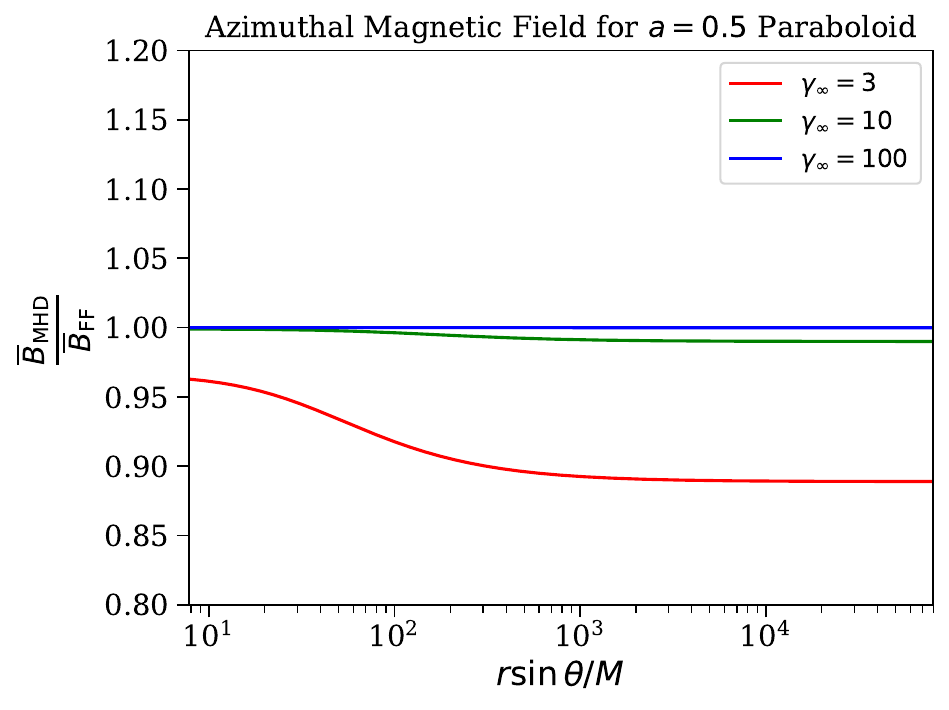}
    \caption{Azimuthal magnetic field for paraboloid ($p=1$) stream function. The force-free azimuthal field $\overline{B}$ is an excellent approximation to the exact MHD field, even when the terminal Lorentz factor is as small as $3$.}
    \label{fig:bphicompare}
\end{figure}

In Figure~\ref{fig:bphicompare}, we plot the paraboloidal azimuthal magnetic field for different values of $\gamma_\infty$ from the GRMHD solutions in the previous section, compared to the FF result. We see that as $\gamma_\infty\to \infty$, the force-free prediction in Eq.~\ref{eq:ffpar} becomes exact, but it is an excellent approximation even for $\gamma_\infty = 3$.

\subsection{Fieldline Rotation Rate $\Omega_F$}
For magnetic fieldlines that thread the black hole horizon and satisfy the force-free constraints in \S\ref{sec:bfieldff}, one can infer a unique value of the fieldline rotation rate $\Omega_F$. In particular, one can switch to a horizon-penetrating coordinate system and demand that the fields remain regular across the horizon, a requirement known as the Znajek condition \citep{znajek_black_1977}. In terms of the variables presented so far, the Znajek condition imposes a relationship between $I$ and $\Omega_F$ as (see Eq.~34 of \citealp{uzdensky_forcefree_2005})\begin{align}
    \Omega_F(\psi)&=\Omega_H+\frac{I\alpha}{4\pi r_+\sin^2\theta \mathcal{B}^r}\bigg|_{r=r_+},
\end{align}
where\begin{align}
    \Omega_H\equiv \frac{a}{2r_+}
\end{align}
is the angular speed of the outer horizon. From this relationship, we can solve for $\Omega_F(\psi)$ by plugging in the poloidal currents (derived from the critical conditions at infinity) in Eqs.~\ref{eq:monoI}-\ref{eq:powerI}.

For the monopole, the result is independent of $\theta$ to leading order in the black hole spin:\begin{align}
\label{eq:omega1}
    \Omega_{F}|_{p=0}&=\frac{a}{8}+\mathcal{O}(a^3),
\end{align}
where we imposed $r_+=2+\mathcal{O}(a)$ throughout. This is the standard result of the BZ monopole \citep{blandford_electromagnetic_1977} and matches the choice of fieldline rotation used in, e.g., \cite{tchekhovskoy_simulations_2008}.
Repeating the identical procedure for $0<p<2$, we find\begin{align}
\label{eq:omega2}
    \Omega_F|_{p>0}&=\frac{a}{4[1+\sec^2(\theta_+/2)]}+\mathcal{O}(a^3),
\end{align}
where $\theta_+$ is the angle at which the fieldline threads the horizon (in the northern hemisphere). Once again, there is a discontinuous change between fieldlines that collimate and fieldlines that do not. To the best of our knowledge, Eq.~\ref{eq:omega2} is a novel result.

As noted earlier, the $p=0$ stream function is an exact solution to the force-free GS equation in Schwarzschild, and is thus a perturbative solution to the GS equation in Kerr at linear order in $a$. Additionally, the $p=1$ stream function can be massaged into an exact solution of the Schwarzschild GS equation by adding an $\mathcal{O}(r^0)$ correction of the form \citep{blandford_electromagnetic_1977,gralla2015note}\begin{align}
\label{eq:bzparaexact}
    \psi|_{p=1,\rm exact}&= r(1-\cos\theta)-2r_+(1-\log 2)\\&\nonumber+r_+(1+\cos\theta)[1-\log(1+\cos\theta)].
\end{align}
Using the Znajek condition, one can derive an analytic expression for $\Omega_F(\psi)$ for this paraboloidal stream function, as in Eqs. G183a-G183c of \cite{chael_black_2023}.

So for all $p$, one can fix a well-defined fieldline rotation rate that produces physical magnetic fields across the black hole horizon. And for the specific cases of $p=0$ and $p=1$, one can write down a stream function that is a perturbative solution to force-free electrodynamics in Kerr.

\subsection{Fluid Velocity}
\label{sec:mhdsec}
Next, a unique prediction for the cold fluid velocity emerges in the force-free limit. As we shall show, this prediction is for the velocity both perpendicular and tangent to the magnetic field.  

In order to satisfy the force-free condition, the velocity perpendicular to the magnetic field must take on a unique value called the drift velocity \citep{mckinney_general_2006,chael_black_2023}:\begin{align}
\label{eq:drift}
  \frac{u^\mu_\perp}{u_\mu n^\mu}&=n^\mu-\epsilon^{\mu\nu\rho\sigma}\frac{n_\nu \mathcal{E}_\rho \mathcal{B}_\sigma}{\mathcal{B}^2}
\end{align} 
where $n_\mu$ is the normal observer defined in Eq.~\ref{eq:normaleq}, and $\mathcal{E}$ and $\mathcal{B}$ are the normal-frame electromagnetic fields defined in Eq.~\ref{eq:emdef}. This is the general relativistic analog of the familiar $\vec{E}\times\vec{B}$ drift from non-relativistic plasma physics. Since all components of $\mathcal{E}^\mu$ and $\mathcal{B}^\mu$ are known in FFE, then the drift velocity can be computed analytically. 

However, the force-free condition does not constrain the component of the four-velocity tangent to the magnetic field. This freedom motivates us to follow \cite{chael_black_2023} and parametrize the force-free  four-velocity as\begin{align}
\label{eq:parameterization}
    u^\mu_{\rm FF}&=\gamma\left(n^\mu+v_{\rm drift}^\mu+\frac{\xi }{\gamma_0}\hat{\mathcal{B}}^\mu\right)
    \equiv
    \gamma\left(n^\mu+v^\mu\right).
\end{align}
Here, $\xi\in (-1,1)$ is an arbitrary boost parameter along the fieldline, $\gamma_0\equiv [1-(v^\mu_{\rm drift})^2]^{-1/2}=(1-\mathcal{E}^2/\mathcal{B}^2)^{-1/2}$ is the Lorentz factor in the case of pure drift ($\xi=0$), $\hat{\mathcal{B}}^\mu\equiv \mathcal{B}^\mu/\sqrt{\mathcal{B}^2}$ is the unit vector tangent to $\mathcal{B}$, and \begin{align}
    v^\mu_{\rm drift}=-\epsilon^{\mu\nu\rho\sigma}\frac{n_\nu \mathcal{E}_\rho \mathcal{B}_\sigma}{\mathcal{B}^2},\qquad \gamma=\frac{1}{\sqrt{1-v^\mu v_\mu}}.
\end{align}
This parameterization automatically enforces that $(u^\mu_{\rm FF})^2=-1$. 

It turns out that a unique value of $\xi$ can be isolated by demanding that the flow conserve energy (assuming we still work in the case of cold plasma). Technically, $E$ and $L$ become formally divergent in the force free limit as $\overline{B}/\eta\to\infty$. However, one can always boost into the co-rotating frame\footnote{Even outside the light cylinders, this procedure is still valid; see Appendix~\ref{sec:appconserve}.}, where $\overline{B}=0$ and hence the locally observed energy becomes finite; in this frame, the winds are driven purely by centrifugal force \citep{spruit_magnetohydrodynamic_1996}. As we show in Appendix~\ref{sec:appconserve}, the energy of the flow in the co-rotating frame is given by\begin{align}
    E_{\rm co}(\psi)&= E-\Omega_FL=-u_t-\Omega_Fu_\phi\\&=\gamma[\alpha-v^\phi(g_{t\phi}+\Omega_Fg_{\phi\phi})],
\end{align}
which (as expected) has no dependence on $\mathcal{B}$. Therefore, the quantity $E-\Omega_FL$ remains a valid conserved quantity if we view force-free electrodynamics as a limit of GRMHD. 

Since the boost parameter $\xi$ is related to the Lorentz factor $\gamma$ via \begin{align}
    \gamma&=\frac{\gamma_0}{\sqrt{1-\xi^2}},
\end{align}
then in order to conserve energy, the following algebraic relation must hold at every point along a fieldline:\begin{align}
   E_{\rm co}= \frac{\alpha\gamma_0-(g_{t\phi}+g_{\phi\phi}\Omega_F)(\gamma_0v^\phi_{\rm drift}+\xi \hat{\mathcal{B}}^\phi)}{\sqrt{1-\xi^2}}.
\end{align}
The solution to this quadratic equation is\begin{align}
\label{eq:parallelboost}
    \xi&=\frac{hf\pm E_{\rm co}\sqrt{E_{\rm co}^2-f^2+h^2}}{E_{\rm co}^2+h^2},
\end{align}
where\begin{align}
    f&=\gamma_0[\alpha-(g_{t\phi}+g_{\phi\phi}\Omega_F)v^\phi_{\rm drift}]
    \\
    h&=(g_{t\phi}+g_{\phi\phi}\Omega_F)\hat{\mathcal{B}}^\phi
\end{align}
The positive root is the correct choice for outflows (inflows) in the northern (southern) hemisphere, and the negative root is the correct choice for inflows (outflows) in the northern (southern) hemisphere. To the best of our knowledge, this result for the field-parallel plasma velocity is new. Note that given Eq.~\ref{eq:parallelboost}, one can view alternative choices of $\xi$ in Eq.~\ref{eq:parameterization} as being due to finite temperature effects.

Unlike the above force-free solution, which is completely analytic, the MHD velocity must be numerically computed as a root of the wind equation. However, a simple and surprisingly accurate approximation for the Lorentz factor in GRMHD simulations is essentially that of FFE with a cap\footnote{In practice, this expression for $\gamma_{\rm MHD}$ can become less than 1 close to the launch point, leading to spacelike solutions. We can account for this issue by taking $\gamma_{\rm MHD}\to (1+\epsilon)\gamma_{\rm MHD}$ for some small $\epsilon$ that ensures the flow is always timelike.}:\begin{align}
\label{eq:gammaapprox}
    \gamma_{\rm MHD}\approx (\gamma_{\rm FF}^{-2}+\gamma_{\rm \infty}^{-2})^{-1/2}.
\end{align}

\begin{figure}[h]
    \centering
    \includegraphics[width=.48\textwidth]{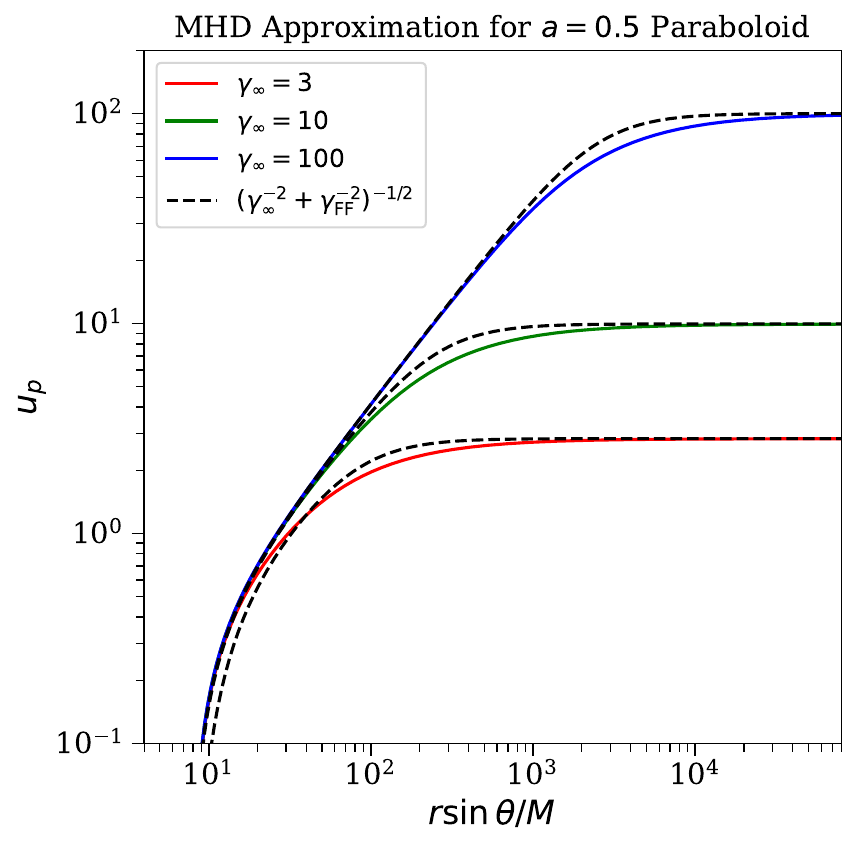}
    \caption{Poloidal four-velocity of wind outflows, shown for exact (in color) and approximate (dashed) solutions to the wind equation. The approximation, which treats MHD as force-free with a maximum Lorentz factor (Eqs.~\ref{eq:gammaapprox}-\ref{eq:mhdu}), works very well.}
    \label{fig:testcap}
\end{figure}

With Eq.~\ref{eq:gammaapprox}, we can derive the components of the approximate MHD four-velocity as\begin{align}
\label{eq:mhdu}
  u^\mu_{\rm MHD}=\gamma_{\rm MHD}(n^\mu+v^\mu_{\rm MHD}) ,\quad v^\mu_{\rm MHD}&=v^\mu_{\rm FF}\sqrt{\frac{1-\gamma_{\rm MHD}^{-2}}{v^\nu_{\rm FF}v_{\nu,\rm FF}}}.
\end{align}
In Figure~\ref{fig:testcap}, we show that this approximation works very well for the sample case of a spin $a=0.5$ black hole in a parabolic field geometry $(p=1)$ that is launched from its stagnation surface.

Thus, Eq.~\ref{eq:gammaapprox} presents a new method to approximately solve the wind equation without the need of a numerical root finder.

\subsection{Summary of the FFE-GRMHD Correspondence}
In this section, we have have shown that deriving force-free electrodynamics as a limit of cold GRMHD with an infinite terminal Lorentz factor leads to unique and analytic results for the force-free drift velocity, including both field-parallel and field-perpendicular components. Using this result, we derived an accurate approximate solution to the MHD wind equation. Combined with a choice of the fluid density profile, this provides a completely analytic description of the plasma inside the jet.



\section{Image Ray Tracing}
\label{sec:imagesec}
In this section, we apply the above formalism to create a model of a jet, which we can ray trace to generate polarized images. Our results build on prior semi-analytic studies of black hole jets (e.g. \citealp{Broderick_2009sil,anantua2020determining,ogihara_radio_2024}).

\subsection{Jet Shape}
To ray trace a jet described by a stream function of the form Eq.~\ref{eq:streamfunc}, we first specify a value of $p$ (the collimation parameter), which fixes the poloidal magnetic field structure. Next, we consider emission from a single value of $\psi$, meaning just one poloidal fieldline\footnote{The emission can in principle be integrated over different values of $\psi$ with an appropriate emissivity model, but we leave that to future work.}; the ``boundary pileup" of plasma onto the jet wall is well motivated by physical models of pressurized outflows and can explain the strong edge-brightening seen in systems like M87 \citep{zakamska_hot_2008}. 

Since the prefactor of the stream function defined in Eq.~\ref{eq:constdefn} can only rescale the total intensity of the image by a global constant, we are free to set $C=\Omega_F^{p-2}\eta\sigma_M=1$, reducing the stream function to\begin{align}
    \psi=r^p(1-\cos\theta).
\end{align}
Unless otherwise specified, we will choose $\psi$ to coincide with the fieldline that intersects the event horizon in the equatorial plane:\begin{align}
\label{eq:psijeteq}
    \psi_{\rm jet}&=r_+^p\left(1-\cos\frac{\pi}{2}\right)=r_+^p.
\end{align}
This ensures that emission close to the black hole comes predominantly from the midplane (consistent with EHT images). The effects of varying $\psi_{\rm jet}$ do not affect our conclusions but are explored in Appendix~\ref{sec:offeqapp}. Since our fiducial choice of fieldline threads the black hole horizon, then we can apply the Znajek condition to determine $\Omega_F$ through Eqs.~\ref{eq:omega1}-\ref{eq:omega2}, which in turn fixes the  azimuthal field structure via Eqs.~\ref{eq:monoI} and \ref{eq:ffpar}.  

\subsection{Plasma Density and Emissivity}
\label{sec:density1}
With the magnetic field structure fixed, we now turn to the dynamics of the plasma inside the jet. To choose a density profile, we will make the standard assumption that the internal energy of the plasma is in equipartition with the magnetic energy stored in the fields \citep{blandford_relativistic_1979,de_young_particle_2006,dexter_size_2012}. While this equipartition assumption is most commonly applied to hot plasma, whose internal energy is dominated by thermal effects, the analogous result for a cold plasma is just that\begin{align}
   \sigma\equiv \frac{|\vec{B}|^2}{\rho}=\text{constant along fieldlines},
\end{align}
where $\vec{B}$ is the magnetic field in the fluid frame. Thus, we have the simple relation $\rho\propto |\vec{B}|^2$, which matches the choice of density profile used in \cite{Broderick_2009sil,takahashi_fast-spinning_2018,ogihara_mechanism_2019}.

Choosing a constant-$\sigma$ density profile means that the continuity equation will not be solved exactly. Instead, it will pick up a source term:\begin{align}
    \nabla_\mu(\rho u^\mu)=S(r,\theta)
\end{align}
for some function $S$.
However, a nonzero source term is to be expected, as plasma will be loaded onto fieldlines due to, e.g., pair production at the stagnation surface or mixing between the higher-magnetization jet and the lower-magnetization disk wind \citep{blandford_electromagnetic_1977,hirotani_pair_1998,hujeirat_model_2003}. Indeed, if one sets $S=0$ and computes the density by solving the sourceless continuity equation, then the density will diverge at the stagnation surface, leading to spikes in the image \citep{kawashima_jet-bases_2021,ogihara_radio_2024}.

To ray trace images of the jet, we also need to endow the plasma with an emissivity function. In particular, we assume that the plasma emits with a power-law electron distribution function (EDF):\begin{align}
    P(\gamma)\propto \gamma^{-s}.
\end{align}
To model synchrotron emission with a spectral index $\alpha_\nu$ (i.e. $I_\nu\propto \nu^{-\alpha_\nu}$), we assign the EDF a power law of $s=1+2\alpha_\nu$. Throughout the rest of this work, we choose $s=2$,
which is well-motivated by theoretical analyses of particle acceleration in plasmas \citep{blandford1978particle,blandford_relativistic_1979}. Additionally, the corresponding spectral index $\alpha_\nu=\frac{1}{2}$ is consistent with the spectral indices of the M87 jet at 24-86 GHz reported by \cite{hada_high-sensitivity_2016}. With this choice of $s$ and $\alpha_\nu$, the Stokes I emissivity function comes out to \citep{Rybicki_Lightman,dexter_public_2016}\begin{align}
\label{eq:emeq}
    j_\nu\propto \rho\nu^{-\alpha_\nu}(|\vec{B}|\sin\theta_B)^{\alpha_\nu+1},
\end{align}
where $\theta_B$ is the fluid-frame angle between the plasma and the magnetic field. This exact choice of emissivity is not critical to our results, which focus instead on polarization orientation.

\begin{figure*}[t]
    \centering
    \includegraphics[width=\textwidth]{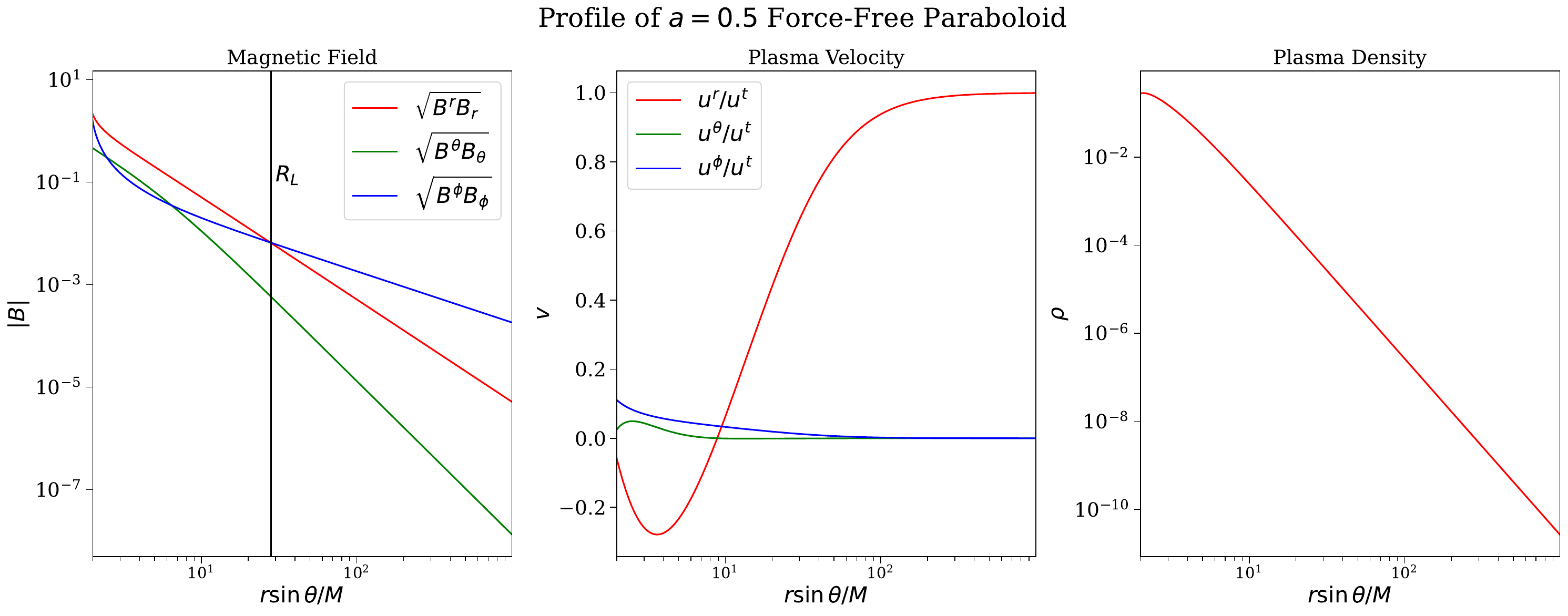}
    \caption{Sample jet profile for force-free, $a=0.5$ paraboloid. The fieldline plotted is $\psi=r_+\log (4)$, which intersects the horizon at the midplane (as one can see by plugging $\theta=\pi/2$ into Eq.~\ref{eq:bzparaexact}). In the left panel, the black vertical line is the light cylinder, which corresponds to a transition from poloidal to azimuthal magnetic field, as well as the transition to relativistic plasma velocities.}
    \label{fig:jetprofile}
\end{figure*}

Altogether, this gives us a model of the magnetic field structure and plasma dynamics in a physical jet that one can ray trace. A sample profile for the equatorial fieldline of the paraboloidal ($p=1$) jet is depicted in Figure~\ref{fig:jetprofile}. Importantly, one can see from this figure that far away from the black hole, the $\phi$ component of the magnetic field dominates. This is characteristic of all jet structures (not just the paraboloid), wherein the field begins to wind up at the light cylinder, as explained in \S\ref{sec:winds}.

\subsection{Polarization}
\label{sec:polarization}
For synchrotron radiation in flat space, the emitted polarization $\vec{f}$ has the important property that it is perpendicular to both the emitted wave-vector $\vec{k}$ and the local magnetic field $\vec{B}$ \citep{Rybicki_Lightman}:\begin{align}
\label{eq:ffluid}
    \vec{f}\propto \vec{k}\times\vec{B}.
\end{align}
In curved space, the polarization 3-vector $\vec{f}$ is then upgraded to a 4-vector $f^\mu$ by requiring that the timelike component of $f^\mu$ be zero in the fluid frame, and that the spacelike components of $f^\mu$ agree with Eq.~\ref{eq:ffluid} in the fluid frame. To transform between the fluid frame (which is locally flat) and the coordinate frame, one can introduce a tetrad $e^\mu_{(b)}$, under which fluid-frame vectors $q^{(a)}$ and coordinate-frame vectors $q^\mu$ are computed as\begin{align}
    q^{(a)}&=\eta^{(a)(b)}e^\mu_{(b)}q_\mu,\qquad q^\mu= e^\mu_{(a)}q^{(a)},
\end{align}
where $\eta^{(a)(b)}$ is the Minkowski metric. More details on the usage of this tetrad are given in \cite{dexter_public_2016}, whose Eqs. 36-39 give the explicit tetrad components.

Equivalently, one can define the polarization in a purely covariant manner \citep{Hou_2024}:\begin{align}
\label{eq:polcompute}
    f^\mu\propto \epsilon^{\mu\nu\rho\sigma}u_\nu k_\rho \mathcal{B}_\sigma,
\end{align}
where $u$ is the four-velocity of the emitter and $k$ is the four-momentum of the emitted photon. Eq.~\ref{eq:polcompute} can be shown to reduce (up to a scalar prefactor) to Eq.~\ref{eq:ffluid} in the fluid frame, where $u=\p_{(t)}$. From Eq.~\ref{eq:polcompute}, one can also see that the polarization will not depend on the component of the four-velocity tangent to $\mathcal{B}$.


\subsection{Radiative Transfer}
The radiative transfer in our model consists of three steps: mapping points on the image to points on the jet, solving the radiative transfer equation inside the jet, and parallel transporting the polarization from the jet to the observer. We outline each of these three steps below. 

\begin{figure}[h]
    \centering
    \includegraphics[width=.4\textwidth]{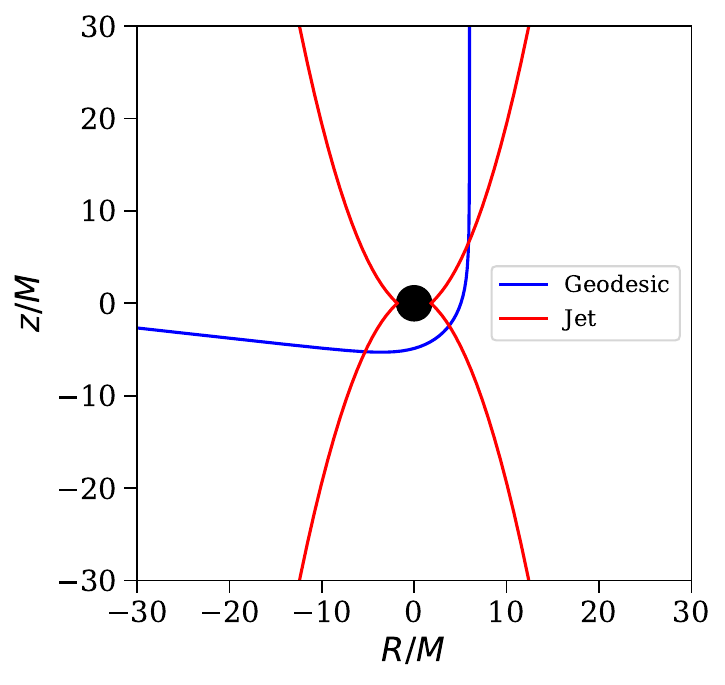}
    \caption{Plot showing trajectory of null geodesic (blue) that arrives on the observer's screen with an impact parameter of $b=10M$. The geodesic crosses the jet (red) three total times before escaping off to infinity. This particular setup was computed for a force-free, $a=0.5$ paraboloid.}
     \label{fig:jetcrossing}
\end{figure}

First, we compute the geodesics that connect the observer to the source. To do so, we specify an array of points in the image plane. Then using the code \texttt{kgeo} \citep{chael_kgeo_2023}, we trace null geodesics from these points towards the hole, keeping track of coordinates $\{r_{\rm geo},\theta_{\rm geo}\}$ at each affine time step. Finally, we perform a Newton-Raphson solve to determine where each geodesic intersects the jet wall, as the intersections correspond to the roots of the equation $\psi_{\rm jet}-\psi(r_{\rm geo},\theta_{\rm geo})$. Multiple roots on a single geodesic frequently occur, as the geodesic typically intersects the forward jet, then the counter-jet, and can loop back around in the case of very strong lensing. A sample geodesic crossing is shown in Figure~\ref{fig:jetcrossing}. In this manner, we can map each point on the image to point(s) on the jet, where the emitted polarization vector is then computed using Eq.~\ref{eq:polcompute}. 

Once we know the emitted polarization vector, we proceed to Step 2: solving the radiative transfer equation inside the emitting medium. We assume the jet is optically thin down to its core, as is expected to be the case for frequencies larger than $\sim 86$ GHz \citep{lee_interferometric_2016,hada_relativistic_2020}. Therefore, the solution to the radiative transfer equation will just be\begin{align}
    I_{\nu,\rm emit}=j_{\nu,\rm emit} \ell_p,
\end{align}
where $\ell_p$ is the length of the geodesic path that traverses the jet wall (as viewed in the emitter frame). The exact expression for $\ell_p$ as a function of fieldline position is derived in Appendix~\ref{app:raytrace}.  

Third, we parallel transport $f^\mu$ from source to observer and project onto the observer's screen, thus creating an image. We neglect Faraday rotation, so this process can be done analytically using the Penrose-Walker constant, which is a quantity derived from $f^\mu$ and $k^\mu$ that is conserved along null geodesics \citep{walker_quadratic_1970,himwich_universal_2020,narayan_polarized_2021,gelles_polarized_2021}. 

After projecting $f$ onto the observer's screen, the result (derived in Appendix~\ref{app:raytrace}) is that the observed specific intensity is given by\begin{align}
\label{eq:iobseq}
I_{\nu,\rm obs}&\propto \frac{g^{2+\alpha_\nu}}{k^{(i)}\hat{w}_{(i)}}\rho R_{\rm emit}\nu^{-\alpha_\nu}(|\vec{B}|\sin\theta_B)^{1+\alpha_\nu},
\end{align}
where $R_{\rm emit}$ is the cylindrical radius of emission, $\hat{w}$ is the unit vector normal to the jet wall, and the Doppler factor $g$ is \citep{narayan_polarized_2021} \begin{align}
    g&=\frac{1}{k^{(t)}}.
\end{align}
The factor of $g^{2+\alpha_\nu}$ is familiar from \cite{blandford_relativistic_1979}.  

With knowledge of the screen-projected polarization and the observed intensity, we will have access to Stokes $I,Q,$ and $U$ at every point on the image.

\section{Analysis of Polarized Images}
\label{sec:imageanalysis}
Using our model, we ray trace sample images for jets with $p=0$, $p=0.75$, and $p=1$, which are displayed in Figure~\ref{fig:polimages}. These images are ray traced at a face-on viewing inclination (see Appendix~\ref{app:incline} for off-axis images), which is appropriate for a low-inclination system like M87 \citep{mertens_kinematics_2016,walker_structure_2018,eht_paper5}. Choosing a face-on inclination also ensures that the images will be axisymmetric, rendering analytic computations tractable.

\begin{figure*}[t]
    \centering
    \includegraphics[width=.32\textwidth]{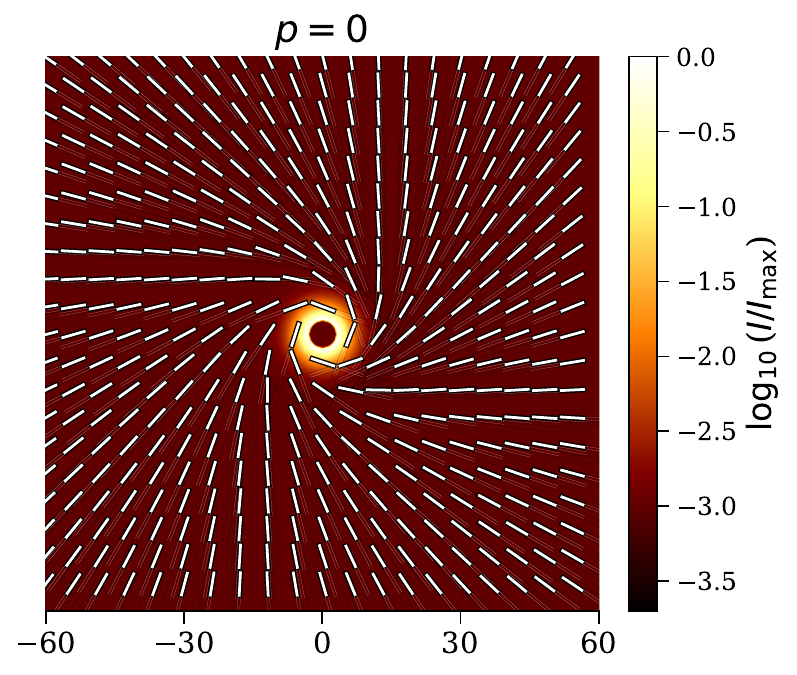}
    \includegraphics[width=.32\textwidth]{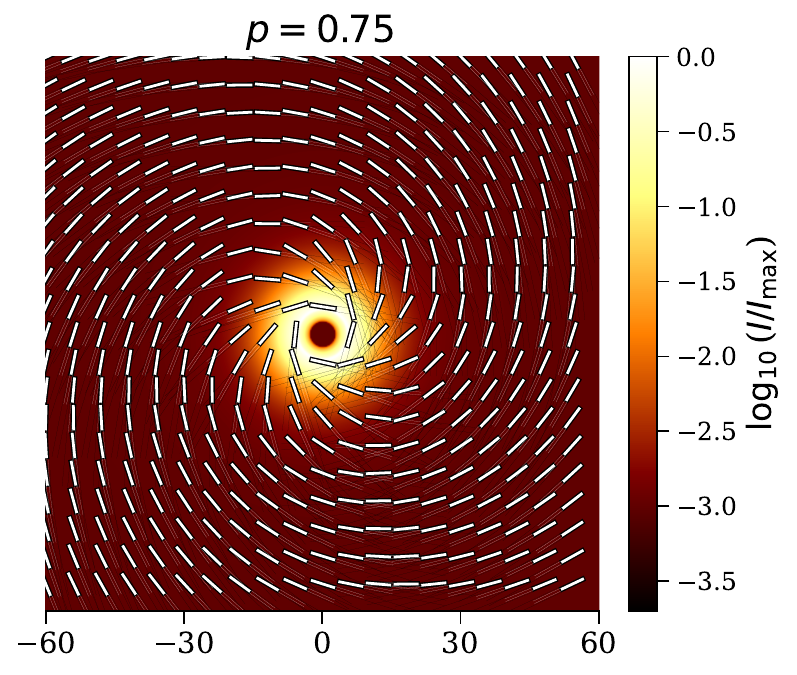}
    \includegraphics[width=.32\textwidth]{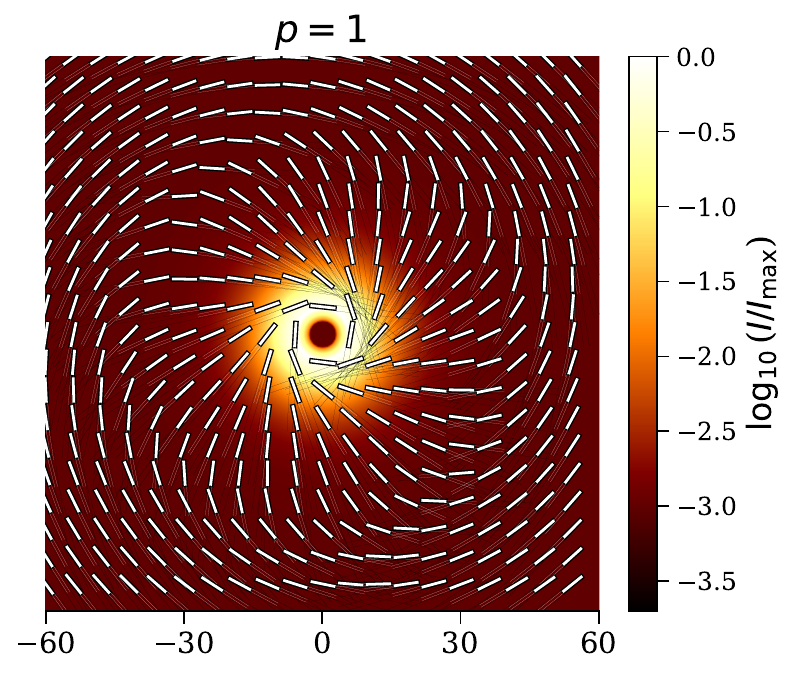}
    \caption{Sample ray-traced images for $a=0.5$ force-free jets at face-on inclination, with the scale plotted in units of $M$ and the ticks tracking the direction of the EVPA. Left is monopole $(p=0)$, middle is $p=0.75$, and right is paraboloid $(p=1)$. The monopolar intensity profile drops the most quickly since its poloidal magnetic field is the weakest ($B^r\sim r^{p-2}$). The abrupt transition in the polarization of the paraboloid is due to relativistic aberration at the light cylinder and is explained in \S\ref{sec:relabb}.}
    \label{fig:polimages}
\end{figure*}
At first glance, the ray-traced image of the paraboloid $(p=1)$ in Figure~\ref{fig:polimages} has several puzzling features.  Naively, one would assume that far from the black hole, the azimuthal magnetic field winds up and dominates over the poloidal field, yielding a radial polarization pattern for face-on observers. But in the image, an azimuthal polarization pattern evidently emerges. Moreover, the transition from radial to azimuthal polarization occurs very abruptly at a well defined location in the image. In this section, we demonstrate that the origin of these polarization swings comes from the interplay between jet and counterjet, as well as relativistic aberration at the light cylinder, as shown in Figure~\ref{fig:annotatefig}.

\subsection{Quantifying Polarization}
To interpret the polarized images, we first need a way of quantifying the polarization pattern. To do so, one can Fourier expand the complex polarization, following \cite{palumbo_discriminating_2020}:\begin{align}
\label{eq:beta2def}
    P(b,\varphi)\equiv Q(b,\varphi)+iU(b,\varphi)=\frac{1}{2\pi}\suml_{m=-\infty}^{\infty}\beta_m(b) e^{im\varphi},
\end{align}
where $b$ is the impact parameter, $\varphi$ is the azimuthal coordinate on the observer's screen, and $Q/U$ are the Stokes parameters. The Fourier coefficients in Eq.~\ref{eq:beta2def} are given by\begin{align}
\label{eq:betadef}
    \beta_m(b)=\int_{0}^{2\pi}d\varphi\, P(b,\varphi)e^{-im\varphi}
\end{align}
and completely specify the polarization pattern at each impact parameter. In the case of a purely axisymmetric image, only the $m=2$ mode is needed. In particular, the complex phase of $\beta_2$, which we will denote $\arg(\beta_2)$, directly encodes the polarization angle at each point $\varphi$. We will adopt the EHT conventions that the observer is placed below the black hole (i.e. at negative $z$), and that the electric vector position angle (EVPA) is measured east of north. With these conventions, the axisymmetric polarization pattern is described by\begin{align}
    \arg(\beta_2)&=-2\,{\rm EVPA}|_{\varphi=\pi/2},
\end{align}
where $\varphi=\pi/2$ corresponds to 12 o'clock on the image. Therefore, $\arg(\beta_2)=0$ corresponds to purely radial polarization, and $\arg(\beta_2)=-\pi$ corresponds to purely azimuthal polarization. 

Our question, then, is why the monopole in Figure~\ref{fig:polimages} has $\arg(\beta_2)\to 0$, while the paraboloid has $\arg(\beta_2)\to -\pi$ at large impact parameter. This discrepancy has to do with the phenomenon of relativistic aberration, as we describe next.

\subsection{Relativistic Aberration}



\label{sec:relabb}
Relativistic aberration --- an apparent change in emission angle due to the emitter's relativistic speed --- is known to play an important role in jet polarization. Specifically, relativistic motion can dramatically alter the synchrotron pitch angle in the plasma frame \citep{blandford_relativistic_1979, konigl_model_1985,pariev_relativistic_2003,lyutikov_polarization_2005,porth_synchrotron_2011}. Here, we build on this work and demonstrate that aberration works to suppress the local azimuthal magnetic field in \emph{all} axisymmetric, highly-magnetized, face-on jets.  

One can show that any solution to force-free electrodynamics must satisfy the asymptotic boundary condition
\citep{armas_consistent_2020}\begin{align}
    \lim_{r\to \infty}I&=-2\pi\Omega_F\sin\theta\p_\theta\psi,
\end{align}
which implies that\begin{align}
   \lim_{r\to\infty} \frac{r\sin\theta \mathcal{B}^\phi}{r\mathcal{E}^\theta}\bigg|_{\rm FF}=1.
\end{align}
In MHD with a finite terminal Lorentz factor, a similar result holds: In Appendix~\ref{app:radiationcondition}, we show that\begin{align}
\label{eq:radcorrection}
    \lim_{r\to\infty}\frac{r\sin\theta \mathcal{B}^\phi}{r \mathcal{E}^\theta}\bigg|_{\rm MHD}=1+\frac{1}{2\gamma_\infty^2}+\mathcal{O}(\gamma_\infty^{-4}).
\end{align}
This equality can be interpreted as a ``radiation condition" \citep{nathanail_black_2014}, where $\mathcal{E} \simeq \mathcal{B}\sim r^{-1}$ so that a finite but nonzero Poynting flux flows in the $\hat{r}$ direction. 

Far from the black hole, the fluid is travelling in the $\hat{r}$ direction close to the speed of light. So letting $\{B^{(r)},B^{(\theta)},B^{(\phi)}\}$, denote the fluid-frame components of the field, we find that a Lorentz boost into the the fluid frame will cause the fields to transform as\footnote{See \cite{cocke1972lorentz,pariev_relativistic_2003} for the Lorentz transformation in terms of the Stokes parameters.}\begin{align}
\label{eq:lorentzboost}
    B^{(r)}&\to B^{(r)}\\\nonumber
     B^{(\theta)}&\to \gamma(B^{(\theta)}+vE^{(\phi)})
     \\\nonumber
     B^{(\phi)}&\to \gamma(B^{(\phi)}-vE^{(\theta)}).
\end{align}
Applying the radiation condition to the transformation of $B^{(\phi)}$, we then have that\begin{align}
\label{eq:radcondition}
    B^{(\phi)}&\to \gamma(r\sin\theta B^\phi-vrE^{\theta})\\&\nonumber=\gamma r E^{\theta}\left(1-v\right)+\mathcal{O}(\gamma_\infty^{-2}).
\end{align}
Thus as $v\to 1$ in a highly magnetized $(\gamma_\infty \gg 1)$ jet, the \emph{local} azimuthal field (which controls the degree of observed radial polarization) vanishes to leading order.

This result extends the calculations of \cite{pariev_relativistic_2003,lyutikov_polarization_2005}, who showed that aberration in cylindrical jets can explain why centimeter-wavelength observations of AGN show either precisely radial or precisely azimuthal polarization patterns (the ``bimodality" problem; \citealp{cawthorne1993milliarcsecond,gabuzda2000analysis,marscher2002high}). Here, we have demonstrated that aberration has a universal effect on \emph{all} highly magnetized jets, so long as they are pointed at the observer. In Appendix~\ref{app:incline}, we show that this effect generalizes to low but nonzero inclinations too; at higher inclinations, different components of the Lorentz boost in Eq.~\ref{eq:lorentzboost} will then contribute to the relevant magnetic field orientation.

Since aberration suppresses the local azimuthal field at leading order, then the 
resultant polarization will depend on the detailed sub-leading physics of the jet model, as we discuss below.

\subsection{Role of Fieldline Collimation}
\label{sec:collimation}
It turns out that the most important property of the sub-leading physics that controls the asymptotic force-free polarization is the \emph{steepness} of the jet's collimation profile. To demonstrate this, we will compute the asymptotic polarization as a function of the collimation parameter $p$ defined in Eq.~\ref{eq:streamfunc}.  

Far from the black hole, we can work in the flat space limit, where the electromagnetic fields in the northern hemisphere become
\begin{align}
\nonumber(\mathcal{B}_r,\mathcal{B}_\theta,\mathcal{B}_\phi)&\to\left(r^{p-2}, -\frac{p\psi}{r\sin\theta},-2\Omega_F \psi\right)\\
   (\mathcal{E}_r,\mathcal{E}_\theta,\mathcal{E}_\phi)&\to \Omega_F\left(-\frac{p\psi}{r},-r^p\sin\theta,0\right),
\end{align}
with $b$ the impact parameter. We can then combine these fields into the flat-space, force-free drift velocity:\begin{align}
    u_{\mu}&\to \gamma_0\left(-1,\frac{\epsilon^{ijk}\mathcal{E}_i\mathcal{B}_k}{\mathcal{B}^2}\right)
    \\
    \nonumber\gamma_0&=(1-\mathcal{E}^2/\mathcal{B}^2)^{-1/2},
\end{align}
where we have disregarded the parallel boost because it cannot affect the polarization. The last ingredient we need to compute the polarization is the wavevector, which in flat space just becomes\begin{align}
        k_\mu&\to (-1,r^{-1}\sqrt{r^2-b^2}, -b,0).
\end{align}
In these expressions, $\theta$ and $b$ can then each be written in terms of $\psi$ and $r$, with the latter following from the flat space ray tracing approximation:\begin{align}
\label{eq:impactparam}
    b\to r\sin\theta = \sqrt{\psi r^{2-2p}(2r^p-\psi)}.
\end{align}
We can then plug all quantities into Eq.~\ref{eq:polcompute} and expand in large $r$, giving:\begin{align}
    f^\theta&\to
        \Omega_F\psi^2 r^{-p}
    \\
    \nonumber f^\phi&\to 
       (p/2-1)\sqrt{2\psi}r^{p/2-1}.
\end{align}
For a face-on viewing geometry, these components are then projected onto the observer's screen to compute $\arg(\beta_2)$.
We find that a precise cancellation of the leading order divergences in the horizontal and vertical components of the polarization occurs at $p=1$, and we obtain\begin{align}
\label{eq:abeq}
   \lim_{b\to\infty} \arg(\beta_2)=\begin{cases}
        0,&p<1
        \\
        \tan^{-1}(2\Omega_F\psi)-\pi,&p=1
        \\
        -\pi,&p>1.
    \end{cases}
\end{align}
The transition at $p=1$ corresponds to the paraboloid. So in the force-free limit, face-on jets with sub-paraboloidal fieldlines will have asymptotically radial polarization, while jets with super-paraboloidal fieldlines will have asymptotically azimuthal polarization. It is the latter class of jets for which relativistic aberration dominates at large radii, forcing $\arg(\beta_2)$ to swing from $0$ to $-\pi$. The paraboloid represents a fringe case, wherein $\arg(\beta_2)$ takes on an intermediate value. 
\begin{figure}[h]
    \centering
    \includegraphics[width=.48\textwidth]{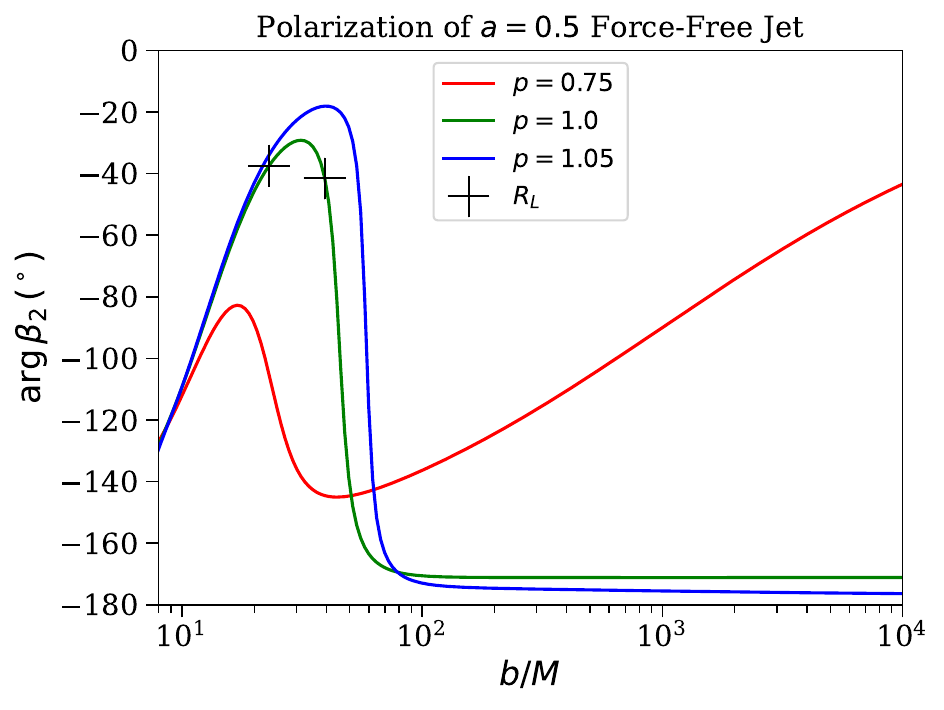}
    \caption{Polarization of face-on, force-free jet as a function of impact parameter $b$, plotted for different values of $p$. The first two lensed images of the light cylinder (jet and counterjet)} are plotted as cross marks and correspond to the swing in polarization.
    \label{fig:pfig}
\end{figure}

This phenomenon is demonstrated in Figure~\ref{fig:pfig}, for which the polarization of the face-on, force-free jet is plotted for three different values of the collimation parameter $p$. The three values respectively correspond to sub-paraboloidal, paraboloidal, and super-paraboloidal fieldlines. In each case, one can see that the polarization undergoes a rapid swing at the light cylinder. At the light cylinder, the magnetic field becomes azimuthal, causing the plasma in the jet to achieve relativistic radial speeds. For $p>1$, the fieldlines are steep enough for aberration to kick in very quickly, causing a sharp swing \emph{down} in $\arg(\beta_2)$ and producing an azimuthal polarization pattern. For $p<1$, the fieldlines are not steep enough for aberration to be important at large radii, so $\arg(\beta_2)$ instead swings \emph{up} and produces a radial polarization pattern.

These findings are broadly consistent with the polarized images produced in \cite{Broderick_2009sil}. In particular, their ``M0" image (high-spin, low-inclination, paraboloidal jet) begins to transition from azimuthal to radial polarization as the near-horizon region is exited. However, it is difficult to make a full comparison to their work, as they ray trace out to just $80\,\mu{\rm as}\sim 20M$, where the plasma is only moderately relativistic ($\gamma\lsim 5$) and so the aberration effects are not captured.

In addition to the role of relativistic aberration in creating a sharp polarization signature, the counterjet also plays an important role, as we explain below.

\subsection{Role of the Counterjet}
\label{sec:cjetsec}
The M87 central engine is known to contain a visible counter-jet at both 43 GHz \citep{ly_attempt_2004,ly_high-frequency_2007,walker_vlba_2008,walker_structure_2018} and 86 GHz \citep{hada_high-sensitivity_2016}. Indeed, semianalytic and GRMHD simulations have confirmed that at near-horizon scales, a combination of gravitational lensing and non-thermal plasma effects can cause the counter-jet to brighten \citep{Broderick_2009sil,dexter_size_2012,moscibrodzka_general_2016,ryan_two-temperature_2018,davelaar_modeling_2019}. These models then predict that outside the near-horizon region, Doppler de-boosting will render the counter-jet far dimmer than its forward counterpart. However, in a highly magnetized jet, the plasma should actually remain non-relativistic all the way out to the light cylinder, meaning that the counter-jet and forward jet can emit with similar intensities out to scales of $\sim 10-100M$. This has profound consequences for the polarized image. 

\begin{figure}[h]
    \centering
    \includegraphics[width=.48\textwidth]{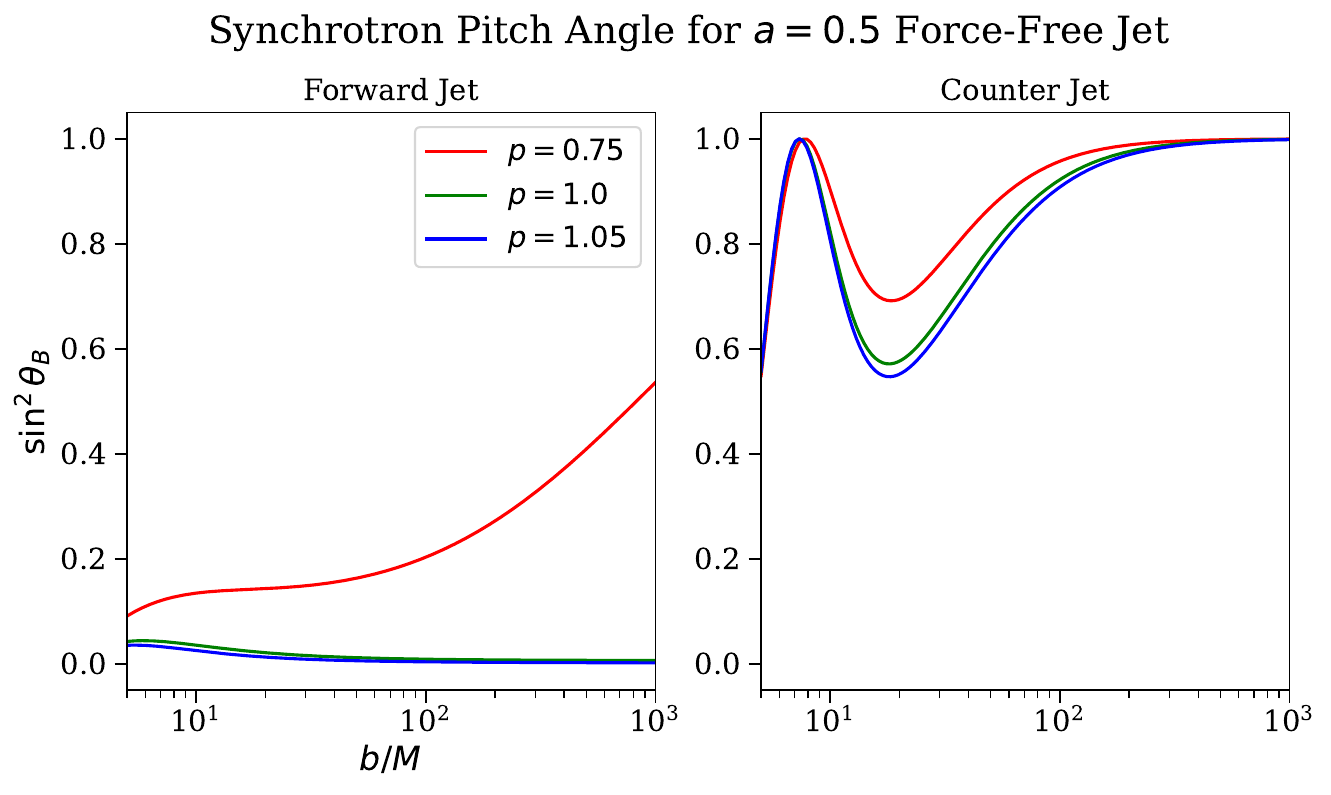}
    \caption{The synchrotron pitch angle $\theta_B$ in both the forward jet and counter-jet as a function of impact parameter. Due to gravitational lensing, the pitch angle is larger in the counter-jet, allowing the counter-jet to be brighter than the forward jet inside the light cylinder.}
    \label{fig:pitchanglefig}
\end{figure}

Indeed, in regions where Doppler effects are negligible, the counter-jet is actually likely to be \emph{brighter} than the forward jet for roughly face-on viewing angles. This is because gravitational lensing will cause the emitted geodesics to bend outward from the counter-jet, thus increasing the pitch angle $\theta_B$ for poloidally dominated fields (which is the case inside the light cylinder). This effect is demonstrated in Figure~\ref{fig:pitchanglefig}, where we decompose the emission into the forward and counter-jet contributions.  

Since the emitted specific intensity is proportional to $\sin^2\theta_B$, then the counter-jet brightness will actually be \emph{enhanced} relative to the forward jet. Notably, this enhacement will occur for any EDF so long as the emissivity depends on a positive power of $\sin\theta_B$\footnote{If the EDF is sufficiently anisotropic with respect to the background magnetic field, it can produce a synchrotron emissivity that scales as a negative power of $\sin\theta_B$ (see, e.g., \citealp{Galishnikova_2023}).}. We note that this lensing effect is different than that of \cite{dexter_size_2012}, which concerns the longer path lengths for geodesics bending around the black hole.

In Figure~\ref{fig:pfig}, the emission is dominated by the counter-jet at small impact parameter, but eventually, the forward jet begins to take over when Doppler effects kick in. Since the paraboloid is strongly collimated, this switch happens precisely at the light cylinder, where the jet becomes  relativistic. Even though the plasma can be moderately relativistic ($\gamma\simeq$ a few) inside the light cylinder, aberration keeps $\theta_B$ small, as the magnetic field will not wind up in the fluid frame. 

For weakly collimated jets $(p<1)$, however, aberration is not a concern; the magnetic field does wind up in the fluid frame, allowing $\theta_B$ to grow far away from the black hole. As a result, the counter-jet can become dimmer than the forward jet well inside the light cylinder, producing an early radial swing in the polarization $\arg(\beta_2)$. This provides additional context for the appearance of the red curve in Figure~\ref{fig:pfig}. At the local maximum in $\arg(\beta_2)$, the counter-jet is turning off, thus allowing the polarization to swing down towards its forward jet value. At the local minimum in $\arg(\beta_2)$, the light cylinder is crossed, meaning $\mathcal{B}_\phi$ becomes dominant in the fluid frame and polarization swings back up to radial.

\subsection{MHD Effects}
\label{sec:mhdeffsec}
Up until now, we have presented polarized images of large-scale jets assuming that the force-free condition holds everywhere. In reality, however, this assumption is likely to break down to some extent due to finite plasma inertia. Here, we examine how to model the impact of this plasma inertia using the approximations summarized in \S\ref{sec:mhdsec}, which are in excellent agreement with our results for full GMRHD winds.

As evidenced by Eq.~\ref{eq:radcondition}, relativistic aberration will stop when $\gamma\simeq\gamma_\infty$ and the force-free approximation breaks down. Past this point, the plasma stops accelerating and the fluid-frame azimuthal field will begin growing again:\begin{align}
    B^{(\phi)}\approx \frac{r\mathcal{E}^\theta}{2\gamma_\infty},
\end{align}
which will eventually diverge. With no aberration to keep $B^{(\phi)}$ small, the polarization will become asymptotically radial, regardless of the fieldline steepness.  

\begin{figure*}[t]
    \centering
    \includegraphics[width=.98\textwidth]{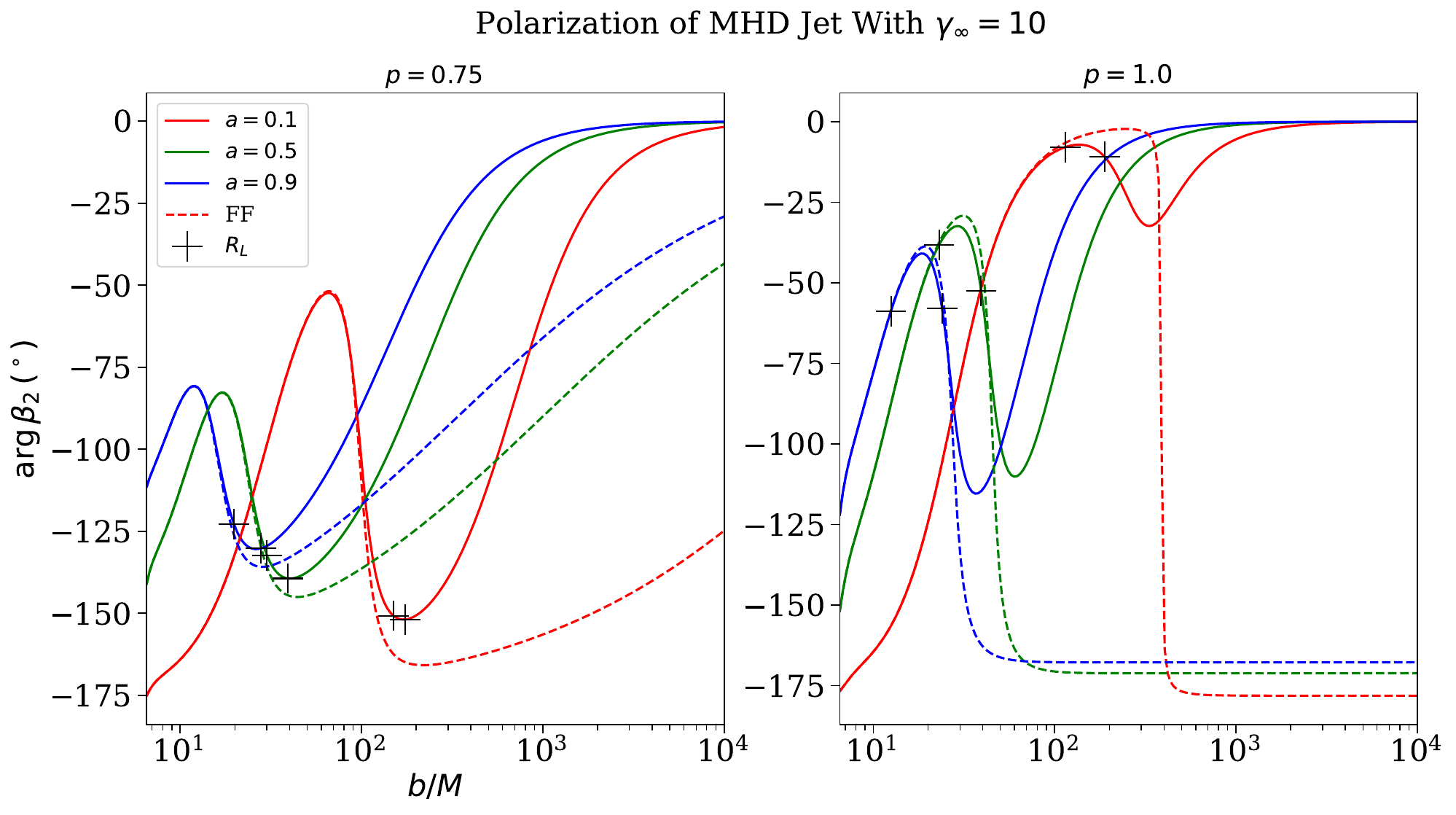}
    \caption{Polarization of $p=0.75$ and $p=1.0$ MHD jets with $\gamma_\infty=10$ and different spins. Like the force-free result (shown as dashed lines), the MHD polarization swings at the light cylinder. But unlike the force-free result, the MHD polarization swings back up when $\gamma\simeq \gamma_\infty$ and aberration becomes subdominant to the increasing field windup.}
    \label{fig:gammacap}
\end{figure*}

While $\gamma_\infty$ can vary dramatically in different astrophysical settings, a typical number in black hole jets is on the order of $\gamma_\infty\sim 10$, as revealed by observations of M87 \citep{biretta_hubble_1999,kino_implications_2022}. The effects of a cap at $\gamma_\infty=10$ are demonstrated in Figure~\ref{fig:gammacap}, in which the polarization is computed for an MHD jet using the approximation of Eq.~\ref{eq:mhdu}.

In weakly collimated jets with $p<1$, the MHD correction is minor; the asymptotic polarization will be radial regardless of $\gamma_\infty$, even though a larger value of $\gamma_\infty$ will cause the polarization to become radial more slowly. In the left panel of Figure~\ref{fig:gammacap}, one can see that the force-free and MHD polarization curves share both the swing at the light cylinder and the radial asymptotic polarization.  

But for strongly collimated jets with $p\geq 1$, the breakdown of FFE can drastically affect the appearance of polarized jet images; in this case, the polarization will swing for a \emph{second} time at the radius where $\gamma\simeq \gamma_\infty$. This effect is shown in the right panel of Figure~\ref{fig:gammacap}, in which one can see that the force-free and MHD polarization curves share the swing at the light cylinder but have different asymptotic values of $\arg(\beta_2)$.

The distinction between FFE and MHD has the potential to dramatically alter the appearance of polarized jet images, especially in cases where the fieldlines are super-paraboloidal. Fortunately, we can use approximations like Eq.~\ref{eq:mhdu} to easily model these effects, and the polarization swings that arise have a particularly clean interpretation in terms of aberration.

\section{Measuring Spin}
\label{sec:spinsec}
In this section, we synthesize our results into a simple strategy for measuring a black hole's spin from polarized observations of its jet on scales of $10M-10^3M$.

As we have shown in \S\ref{sec:collimation}, the windup of the magnetic field leaves a well-defined imprint as a spatial polarization swing at the light cylinder. In strongly collimated jets $(p>1)$, relativistic aberration keeps the fluid-frame azimuthal field small, forcing $\arg(\beta_2)$ to swing to $-\pi$. In weakly collimated jets $(p<1)$, aberration is insufficient to suppress the azimuthal field, causing $\arg(\beta_2)$ to swing to $0$. The relative brightness of the counter-jet at the light cylinder then controls the direction of the swing. These results can be summarized by:\begin{align}
\label{eq:swing1}
    \frac{d\,\arg\beta_2}{db}\bigg|_{R\approx R_L}&=0
    \\
    \label{eq:swing2}
    {\rm sign}\left(\frac{d^2\arg\beta_2}{db^2}\right)\bigg|_{R\approx R_L}&=\begin{cases}
        +1,&p<1
        \\
        -1,&p\geq 1.
    \end{cases}
\end{align}
Eqs.~\ref{eq:swing1}-\ref{eq:swing2} are observable features of the jet that directly encode the light cylinder's location. And with knowledge of the light cylinder's location, one can obtain the black hole's spin, as $R_L\sim\Omega_F^{-1}\sim a^{-1}$. This spin dependence of the polarization swing can be seen in Figure~\ref{fig:gammacap}: increasing the black hole's spin causes the polarization curves to translate to the left as the light cylinder moves in. Thus, one can measure a black hole's spin by searching for a polarization swing (i.e. an extremum of $\arg\beta_2$) and identifying it with the light cylinder.  

Up until now, most attempts to measure supermassive black hole spins have been performed via X-ray reflection spectroscopy, wherein the broadening of iron emission lines can in principle be used to identify the innermost stable circular orbit (ISCO) and hence the spin of the black hole \citep{fabian1989x,george1991x,brenneman2006constraining,reynolds2008broad}. This technique is not applicable to the low-luminosity AGN that are believed to be most common throughout the universe, as their accretion rates are not high enough to produce the characteristic X-ray spectra \citep{reynolds2021observational,ricarte2022ngeht}. Our method of measuring spin, on the other hand, is entirely applicable to these AGN due to the observability of their large-scale jets.

Our proposed method to measure spin through polarized observations of the light cylinder is powerful for two primary reasons:  

\textbf{First, this method has the potential to be robust to many astrophysical uncertainties.} As Figure~\ref{fig:gammacap} demonstrates, a polarization swing at the light cylinder occurs regardless of the terminal Lorentz factor and regardless of the fieldline collimation profile. In fact, we expect this signature to persist even if we were to abandon our idealized jet model in favor of one that incorporates more astrophysically realistic jets (e.g. finite jet wall thickness, different electron distribution function, non-axisymmetry). Poynting-dominated jets from black holes will generically transition from poloidally-dominated to azimuthally-dominated fields near the light cylinder, so the observed polarization should generically change there. Future work to explicitly test this prediction using time-dependent simulations will be very valuable.  

One potentially important complication is Faraday rotation: AGN (particularly BL Lac objects and quasars) tend to have strong rotation measure gradients arising from powerful helical magnetic fields \citep{Asada_2002,gabuzda_helical,broderick_signatures_2009,broderick_parsec-scale_2010,gomez2012helical,hovatta_mojave_2012}. However, external Faraday rotation will serve only to shift the endpoint values of $\arg(\beta_2)$ in the polarization swing; the existence of the swing itself will not change. Internal Faraday rotation may pose more of a challenge to model, but it will likely contribute less at 230 GHz compared to the centimeter-wavelength observations referenced above. In any case, multi-frequency observations will be crucial to accurately account for and mitigate Faraday effects in jet polarimetry.

\textbf{Second, this method is particularly diagnostic in the low-spin regime.} Since $R_L\sim a^{-1}$, the location of the polarization swing will be extremely sensitive to spin when $a\ll 1$. Indeed, one can see that the location of the $a=0.1$ polarization swing in Figure~\ref{fig:gammacap} occurs at an impact parameter $\sim 10$ times larger than that of the $a=0.9$ swing. Thus, one does not need unreasonably high spatial resolution to determine whether a black hole is high-spin $(a\gtrsim 0.1)$ or low-spin $(a\lsim 0.1)$; pinning down an approximate location of the polarization swing will be sufficient. This method serves as an important complement to proposals to measure spin through the shape of the strongly lensed photon ring, which is most deformed from circularity at high spin \citep{bardeen_1973,johnson2020universal,gralla2020observable}. 

In assessing the role of spin in producing radial variation in jet polarization, it is important to account for the fact that other physical processes are also capable of causing qualitatively similar polarization swings; one must take caution to ensure these do not get confused with the light cylinder. For example, the dimming of the counter-jet (described in \S\ref{sec:cjetsec}) and the breakdown of the force-free approximation (described in \S\ref{sec:mhdeffsec}) can both cause the polarization to vary as a function of distance from the black hole. Localized changes in jet inclination \citep{lyutikov_polarization_2005} and plasma shocks \citep{denn2000very} can cause large polarization changes as well. Thus, it will be important to map out the jet polarization over a range of radii to be confident in identifying the light cylinder. Likewise, it will be important to combine our method for measuring spin with other astrophysical measurements of the jet. In particular, total intensity observations of AGN can be used to compute model-agnostic values of $p$, $\gamma_\infty$, and counter-jet brightness ratios (see, e.g. \citealp{asada_structure_2012,park2019kinematics,kino_implications_2022}), thus breaking the degeneracies raised above. Finally, near-horizon observations of polarization can also break degeneracies and point to model-agnostic predictions of spin, as detailed in \cite{Hou_2024}.

Indeed, we expect the character of the jet's intensity profile (intensity as a function of impact parameter) to change at the light cylinder too. Inside the light cylinder, the fields sourcing the polarized emission scale as $B\approx B_p\sim r^{p-2}$, which differ from the fields sourcing the polarized emission outside the light cylinder (where $B\approx B_\phi\approx r^{-1}$). Furthermore, the plasma is non-relativistic inside the light cylinder with constant Doppler factor $g\approx 1$, whereas $g$ grows with radius as the plasma accelerates outside the light cylinder. Therefore, we expect the intensity to drop more steeply inside the light cylinder than it does outside the light cylinder, and we present a detailed derivation of this result in Appendix~\ref{app:raytrace}. In this manner, one can confirm the existence of the light cylinder from a polarization swing by identifying a subsequent break in the intensity power law. This result also explains why the images in Fig.~\ref{fig:polimages} become so dim outside $\sim 20M$: the steep drop in intensity interior to the light cylinder renders the outskirts of the images very faint.

Accurate constraints on black hole spins are important for a multitude of reasons. Measurements of spin go hand-in-hand with measurements of mass to characterize strophysical black holes, which are believed to be described by these two numbers alone.\footnote{The no-hair conjecture allows for black holes to accrue electric charge as well, though this likely does not affect astrophysical black holes, which accrete in charge-neutral environments.} Moreover, measurements of black hole spin can provide important information about merger histories, jet spindown, and cosmological evolution of black holes, enabling a better understanding of black hole growth and feedback \citep{ricarte2022ngeht,narayan_jets_2022,ricarte2023recipes}. Spin measurements will also play a key role in establishing the BZ mechanism as the origin of astrophysical jets, as the BZ jet power is a strong function of black hole spin \citep{blandford_electromagnetic_1977}.

Finally, it is important to combine various methods of measuring spin so that they can serve as consistency checks on each other. Combining polarized observations of the jet with observations of the photon ring, for example, will allow for robust inference of black hole spin, while simultaneously shedding new light on the ways in which black holes launch jets from the near-horizon region to large spatial distances. 

\section{Summary and Discussion}
\label{sec:discussion}
We have developed a semi-analytic, physically motivated model of astrophysical jets that allows us to predict the polarization for a range of magnetic field geometries and plasma properties. The model inputs include a collimation parameter $p$, which controls the steepness of the magnetic fieldlines via Eq.~\ref{eq:streamfunc} (see Fig. \ref{fig:fieldlineshape}), as well as a terminal Lorentz factor $\gamma_\infty$, which controls the plasma velocity asymptotically far from the black hole. In this paper, we have focused on the polarization of roughly face-on jets (as is the case in M87), but the extension to arbitrary viewing angles is conceptually straightforward.   

\begin{figure*}[t]
    \centering
    \includegraphics[width=.8\textwidth]{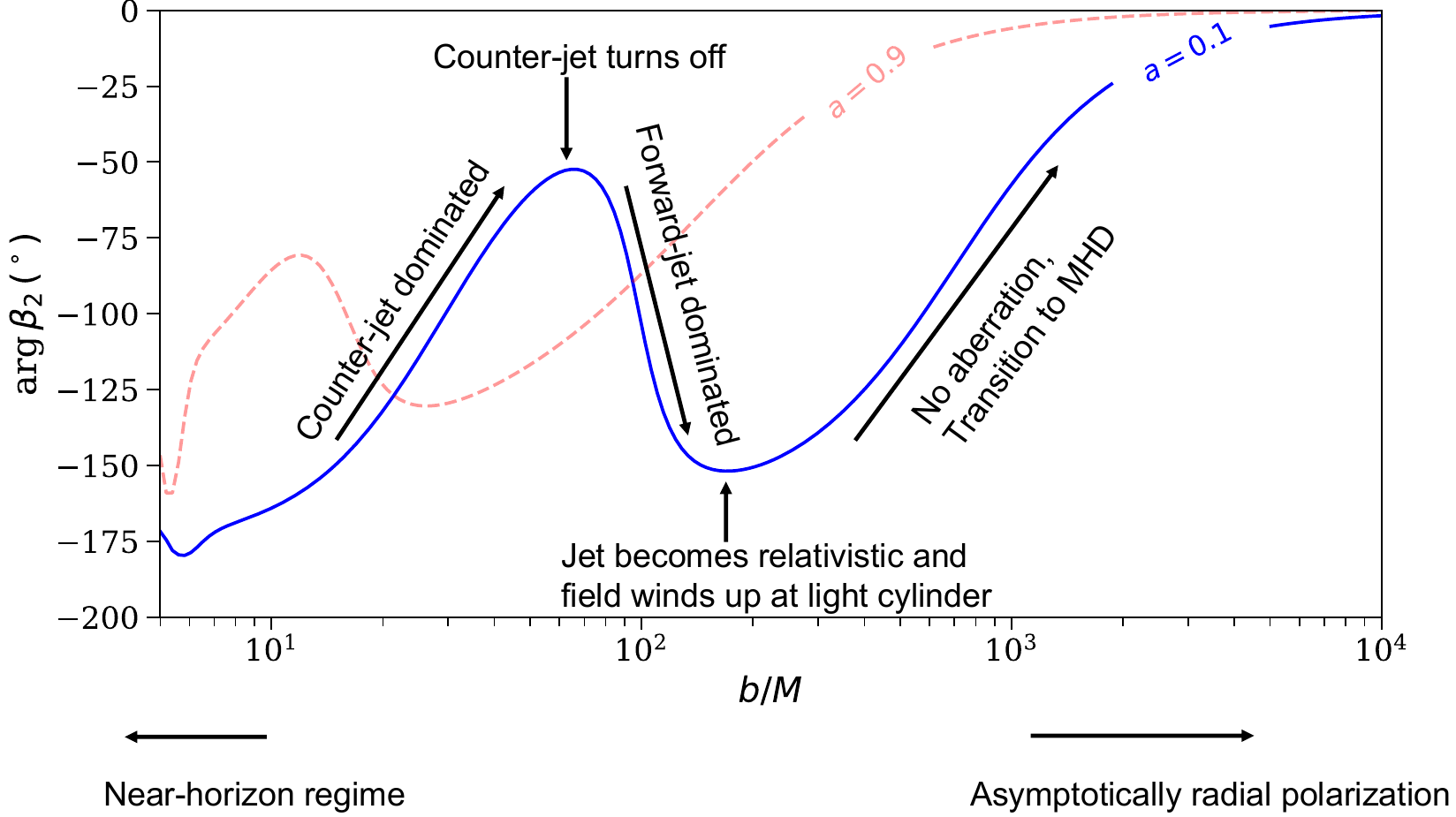}
    \caption{Polarization of a face-on jet as a function of impact parameter, plotted for $p=0.75$ and $\gamma_\infty=10$. Sharp variations occur at characteristic locations in the flow, offering a strong observational signature of spin; the solid blue line corresponds to $a=0.1$ and has polarization swings that lie significantly farther out than the dashed red line, which corresponds to $a=0.9$. The significance of the parameters $\gamma_\infty$ and $p$, as well as the nature of each swing, is explained in \S\ref{sec:imageanalysis}.  Extensions to a finite viewing angle relative to the jet axis appropriate to M87 are given in Figs. \ref{fig:incimage}/\ref{fig:incimage2} of Appendix \ref{app:incline}, and are similar to the results shown here.}
    \label{fig:annotatefig}
\end{figure*}

Using our model, we have shown that the polarization of the jet is expected to undergo large radial swings in up to three places in the image:
\begin{enumerate}
    \item The radius where the counter-jet turns off due to Doppler effects
    \item The light cylinder, where the coordinate-frame magnetic field winds up and the jet becomes relativistic
    \item The radius where the force-free approximation breaks down because of plasma inertia, causing the outflow to reach its terminal Lorentz factor
\end{enumerate}
At each of these locations, the polarization is expected to change rapidly as a function of radius, thus producing a highly observable signature that can be quantified in terms of the parameter  $\arg\beta_2$ (defined in Eq.~\ref{eq:beta2def}). The combination of these effects is summarized visually in Figure~\ref{fig:annotatefig}, and we briefly expand upon each of these effects below.

Inside the light cylinder, the plasma is non-relativistic, meaning that the counter-jet can be seen. In fact, for roughly face-on jets, the counter-jet may be brighter than the forward jet, as gravitational lensing causes the synchrotron pitch angle $\theta_B$ (the angle of the emitted photon relative to the magnetic field) to be larger for emission behind the black hole (see Figure~\ref{fig:pitchanglefig}); this enhances the flux of the counter-jet relative to the forward-jet. However, at the radius where the counter-jet eventually dims due to Doppler effects, the polarization will swing to track the forward jet instead.  This produces the first polarization swing as a function of radius in Figure \ref{fig:annotatefig} at $\sim 50 M$ for $a = 0.1$. 

Outside the light cylinder, the polarization will swing again as the magnetic field starts to wind up to become toroidally dominated.  This produces the polarization swing as a function of radius at $\sim 200M$ for $a = 0.1$ in Figure \ref{fig:annotatefig}.   The precise nature of this polarization swing depends on the fieldline collimation $p$ via Eq.~\ref{eq:abeq}; at the light cylinder, sub-paraboloidal fieldlines ($p<1$) will begin to produce radial polarization (as in Fig. \ref{fig:annotatefig}), whereas super-paraboloidal fieldlines $(p>1$) will begin to produce azimuthal polarization. This distinction, which arises due to the effects of relativistic aberration on the synchrotron pitch angle, can be seen clearly in Figure~\ref{fig:pfig}. 

Far from the black hole, the force-free approximation breaks down, halting the growth of the drift velocity (Eq.~\ref{eq:drift}) as the plasma reaches its terminal Lorentz factor $\gamma_\infty$. This could induce a third potential polarization swing, as the lack of plasma acceleration means that aberration can no longer keep up with the field windup. This phenomenon is seen in Figure~\ref{fig:gammacap}.

The light cylinder --- where the magnetic field transitions from poloidally dominated to azimuthally dominated --- is a dynamically important surface that serves as a common thread among each of the effects described above. Indeed, the polarization swing at the light cylinder \emph{always} occurs (regardless of choice of $p$, $\gamma_\infty$, etc.), offering a unique and robust imprint of black hole spin as discussed in depth in the previous section. An observed break in the total intensity profile can serve as additional evidence of the light cylinder too.

In computing the polarization patterns of black hole jets, we have shown that cold GRMHD models of stationary, axisymmetric jets can be very well-approximated using the analytically simpler formalism of force-free electrodynamics. This is possible due to two novel results about the plasma dynamics: that cold plasma \emph{does} have a unique field-parallel boost velocity in the force-free limit (Eq.~\ref{eq:parallelboost}), and that MHD corrections at large distances from the black hole can be modelled extremely well by simply capping the force-free Lorentz factor in a smooth way (Eq.~\ref{eq:gammaapprox}, Figure~\ref{fig:testcap}).

The formalism developed here can play an important role in motivating and interpreting future VLBI campaigns of black hole jets; the next round of interferometric array upgrades will enable polarized observations with unprecedented dynamic range, potentially revealing the signatures of the light cylinder and black hole spin predicted here.  While we expect the polarization swing at the light cylinder to be a robust signature, there are still important extensions of our model needed to reflect more realistic models of astrophysical jets. 

First, we hope to systematically investigate the effects of observer inclination. While Appendix~\ref{app:incline} demonstrates that the spin dependence of the polarization swings still holds for small but nonzero inclination angles, we do not know how to best interpret the polarization when the jet is viewed closer to edge-on. The parameter $\arg(\beta_2)$ becomes hard to interpret for edge-on images, as it no longer characterizes the complete Fourier-domain polarization pattern (Eq.~\ref{eq:beta2def}). Indeed, even for $i= 17^\circ$, the effects of finite viewing angle strongly break the axisymmetry far from the hole. Since the light cylinder always represents a transition from poloidal to azimuthal magnetic field, we expect \emph{some} kind of polarization swing to happen there regardless of the viewing geometry. But the specifics still need to be worked out. Indeed, at nonzero viewing inclinations, it will be important to incorporate the effects of limb-brightening to match onto observations \citep{reid1989subluminal,kovalev2007inner,kim2018limb,lu2023ring}. To model this effect, a different emissivity model will be needed.

Second, we hope to model the effects of emission from multiple fieldlines. Indeed, including emission interior to the jet wall is particularly important, as aberration can cause the edge (sheath) of the jet to have a different polarization than the interior (spine) of the jet \citep{gabuzda_helical,lyutikov_polarization_2005,porth_synchrotron_2011,clausen2011signatures,gabuzda2014spine}. This may cause the observed polarization swings to be smaller in magnitude than those predicted by our semi-analytic model. In any case, different fieldlines should produce similar qualitative polarization patterns in low-inclination images, as discussed in Appendix~\ref{sec:offeqapp}. Additionally, we need to model differences in polarization fraction at different locations of the jet. In M87, the jet polarization fraction has been measured to be relatively low ($3-20\%$; \citealp{hada_high-sensitivity_2016}).

Ultimately, we plan to check if our proposed polarization signatures emerge in numerical GRMHD simulations of relativistic jets from spinning black holes. We have varied all the parameters of our model's underlying physics (including terminal Lorentz factor $\gamma_\infty$, collimation parameter $p$, EDF power law $s$, and the scaling of the jet width $W$), and the spin-dependent polarization swings remain throughout. However, GRMHD simulations will capture effects like mixing between disk and wind, as well as time-dependent and non-axisymmetric structures that cannot be modelled in our analytic framework at all. Because synchrotron emission is non-linear in the plasma properties, it is possible that the time-averaged magnetic field inferred through polarization is not the same as the time averaged magnetic-field itself; this can be assessed using time-dependent GRMHD simulations with models for non-thermal electron emission. 

One particularly interesting feature that will be captured in numerical simulations is the kink instability, wherein the high magnetic pressure from toroidal fields can cause the magnetic field to unwind \citep{shafranov1956stability,kruskal1958instability,begelman_1998_jets,lyubarskii1999kink,moll2008kink}. One might worry that as a result of the kink instability, the ordered toroidal fields in our analytic model may not be realistic exterior to the light cylinder. Fortunately, past numerical GRMHD simulations find that the kink instability likely does \emph{not} play a dynamical role across large swaths of the jet \citep{mckinney_general_2006}. The instability is most likely to affect the field structure when the jet re-collimates (i.e. becomes conical) at large radii, e.g., at the Bondi radius \citep{tchekhovskoy2016three,bromberg2019kink}. For typical AGN in massive galaxies, this re-collimation happens at $10^5-10^6M$ \citep{asada_structure_2012,nakamura2018parabolic,kovalev2020transition}, which lies well beyond the light cylinder. We expect that the polarization signature we have identified, therefore, should remain unchanged. But it will still be crucial to test this hypothesis using  simulations. 

As we improve our model and test our predictions against numerical simulations, we will continue to focus on the importance of measuring spin. The polarization swing at the light cylinder is a clear, robust, and highly diagnostic signature that will be observable by future arrays like the ngEHT and future space-VLBI instruments like BHEX. As interferometric upgrades come, such observational signatures can play an important role in performing the most constraining black hole spin measurements to date, in precisely the systems where spin-energy is believed to power relativistic jets.\medskip 

\noindent We thank Charles Gammie, Michael Johnson, Alexandru Lupsasca, Ramesh Narayan, Hung-Yi Pu, Daniel Palumbo, Frans Pretorius, Paul Tiede, and George Wong for helpful discussions. ZG was supported by a National Science Foundation Graduate Research Fellowship. AC was supported by the Princeton Gravity Initiative. This work was supported in part by a Simons Investigator Award to EQ and benefited from EQ's stay at the Aspen Center for Physics, which is supported by National Science Foundation grant PHY-2210452.

\appendix

\section{Features of Stationary, Axisymmetric, Cold GRMHD}
\subsection{Field Conventions}
\label{sec:conventions}
In \S\ref{sec:setup}, we define the electromagnetic fields from the normal observer's frame. Here, we present two alternative conventions for defining the fields in curved space. First, several works in the GRMHD literature (e.g. \citealp{komissarov_godunov-type_1999,gammie_harm_2003}) define the primitive electromagnetic fields directly from the ``coordinate-frame":\begin{align}
\label{eq:defb2}
    E^\mu&=F^{t\mu},\quad B^\mu=-(\star F)^{t\mu}.
\end{align}
In practice, this choice just rescales the ZAMO-frame fields used in this paper by a factor of the lapse:\begin{align}
    \frac{\mathcal{E}^\mu}{E^\mu}\,(\text{no sum})=\frac{\mathcal{B}^\mu}{B^\mu}\,(\text{no sum})=\alpha.
\end{align}
Since the four-velocity of the fluid always depends on a ratio of electromagnetic fields, this discrepancy will drop out in computing the fluid dynamics. 

Second, some authors (e.g. \citealp{camenzind_hydromagnetic_1986,takahashi_magnetohydrodynamic_1990,takahashi_constraints_2008,pu_properties_2020}) define the magnetic field from the \emph{lowered} Faraday dual:\begin{align}
    \mathbf{B}_\mu&\equiv (\star F)_{t\mu},
\end{align}
which would give a different result than naively lowering the definition of $B^\phi$ from Eq.~\ref{eq:defb2}:\begin{align}
    \mathbf{B}_\mu\neq g_{\mu\nu}(\star F)^{t\nu}.
\end{align}
Indeed, $\mathbf{B}_\phi$ is the same quantity as the variable $\overline{B}$ that we defined in Eq.~\ref{eq:bbareq}. In this convention, the fields can be understood as those viewed in the frame of the timelike killing vector $K^\mu=(1,0,0,0)$, which does not generically correspond to a physical four-velocity \citep{chael_black_2023_v2}. 

These various constructions of the magnetic field agree in Minkowski space, but one must be explicit about choosing one of the three schemes in curved spacetime.
\subsection{Conserved Quantities}
\label{sec:appconserve}
Here, we review the derivation of energy and momentum conservation presented in Eq.~\ref{eq:etaeq}, and we provide a physical interpretation of their difference $E-\Omega_FL$ that is used to derive the parallel boost in \S\ref{sec:mhdsec}. For a derivation of the other conserved quantities $\eta$ and $\Omega_F$, see \cite{phinney_theory_1984}.



To derive $E$ and $L$, we turn to the stress tensor, which is conserved in the sense that $\nabla_\mu T^{\mu\nu}_{\rm MHD}=0$. From this conservation law, one obtains two Noether currents $j^\mu_t=-K_\nu T^{\mu\nu}$ and $j^\mu_\phi=R_\nu T^{\mu\nu}$, whose existence follows from the fact that $K^\mu=(1,0,0,0)$ and $R^\mu=(0,0,0,1)$ are killing vectors of the Kerr spacetime. Letting $A\in\{r,\theta\}$ be a poloidal index, we derive the poloidal components of these currents as\begin{align*}
    j^A_t&=-K^\nu T^A_\nu=-F^{A\mu}F_{t\mu}-\rho u^Au_t=\rho u^A\left[-u_t-\frac{\overline{B}\Omega_F}{\eta}\right]
    \\
    j^A_\phi&=R^\nu T^A_\nu=F^{A\mu}F_{ \phi\mu}+\rho u^Au_\phi=\rho u^A\left[u_\phi-\frac{\overline{B}}{\eta}\right],
\end{align*}
where the mass-loading $\eta$ is defined in Eq.~\ref{eq:etaeq}.
This motivates our definition of specific energy and specific angular momentum as\begin{align}
    E\equiv \frac{j^A_t}{\rho u^A}=\frac{\alpha j_t^A}{\eta \mathcal{B}^A}=-u_t-\frac{\overline{B}\Omega_F}{\eta},\qquad L\equiv \frac{j^A_\phi}{\rho u^A}=\frac{\alpha j^A_\phi}{\eta \mathcal{B}^A}=u_\phi-\frac{\overline{B}}{\eta}.
\end{align}
Then $E$ is conserved along fieldlines, as \begin{align}
   \mathcal{B}^\mu \nabla_\mu E\propto  \eta \alpha^{-1}\mathcal{B}^\mu \nabla_\mu E =\nabla_\mu (\eta\alpha^{-1}\mathcal{B}^\mu E)=\nabla_A(\rho u^A E)=\nabla_\mu j_t^\mu=0,
\end{align}
where we employed the no-monopole constraint $\nabla_\mu (\star F)^{t\mu}=\nabla_\mu(\alpha^{-1}\mathcal{B}^\mu)=0$.
An identical calculation with $j^A_t\to j^A_\phi$ shows that $L$ is conserved too. 

Now consider an observer who is co-rotating with the fieldlines. To this observer, time-translations are given by the four-vector $\zeta^\mu\equiv (1,0,0,\Omega_F)$. Therefore, the poloidal momentum density, as measured in the co-rotating frame, is given by\begin{align}
    -\zeta^\mu T^A_\mu&=(E-\Omega_FL)\rho u^A,
\end{align}
which does not depend on the magnetic field at all. This makes sense since there are no magnetic forces in this frame. This expression motivates us to define a ``co-rotating energy" as\begin{align}
    E_{\rm co}\equiv E-\Omega_FL=-u_t-\Omega_Fu_\phi,
\end{align}
which is the general relativistic generalization of what \cite{spruit_magnetohydrodynamic_1996} calls ``$E$". For more intuition about the meaning of $E_{\rm co}$, we can Taylor expand in the Newtonian, special relativistic limit, where $u_t=-\gamma$ and $u_\phi=\gamma r^2\sin^2\theta \Omega_P$, giving\begin{align}
    E_{\rm co}\to 1+\frac{1}{2}[v_p^2+r^2\sin^2\theta(\Omega_P-\Omega_F)^2]-\frac{1}{2}r^2\sin^2\theta\Omega_F^2+\mathcal{O}(v^3),
\end{align}
with $\Omega_P=d\phi/dt$ the angular speed of the plasma. This is precisely the sum of rest mass, kinetic, and centrifugal terms in the co-rotating frame. 

We note that the co-rotating frame does not always correspond to a physical observer, as the frame can be spacelike very far from or very close to the black hole. But the quantity $E_{\rm co}$ remains conserved regardless.

\subsection{The Wind Equation}
\label{sec:windeqn}
Here, we provide an abbreviated  derivation of the wind equation introduced in \S\ref{sec:winds}, referring the reader to \cite{phinney_theory_1984} for more details. We begin with the $\theta$ component of the MHD condition, which fixes $\overline{B}$ in terms of the four-velocity:\begin{align}
\label{eq:bbarsolve}
    \nonumber 0&=u^\mu F_{\mu \theta}\propto \mathcal{B}^r(u^\phi-\Omega_Fu^t)-\mathcal{B}^\phi u^r
    \\
    \Longrightarrow\overline{B}&=\Delta \alpha^{-1}\sin^2\theta \mathcal{B}^r\left(\frac{u^\phi-\Omega_F u^t}{u^r}\right).
\end{align}
Regularity at the launch point (where $u^r=0$) thus implies that\begin{align}
\label{eq:launcheq2}
    \frac{u^\phi}{u^t}\bigg|_{R=R_0}=\Omega_F,
\end{align}
which is used in Eq.~\ref{eq:launcheq}. Furthermore, this expression for $\overline{B}$ can then be plugged into the conserved quantities so that $u^t$ and $u^\phi$ are explicitly given as functions of $E$, $L$, and $u^r$:\begin{align}
\label{eq:usolve}
u^t&=\frac{\Delta\sin^2\theta(E-\Omega_FL)-\frac{\alpha\eta u^r}{\mathcal{B}^r}  (Eg_{\phi\phi} +Lg_{t\phi}) }
{N_{\rm co}\Delta\sin^2\theta-\frac{\alpha\eta u^r}{\mathcal{B}^r}\left(g_{t\phi}^2-g_{\phi\phi} g_{tt}\right)}
,\quad 
u^\phi=\frac{\Omega_F\Delta\sin^2\theta(E-\Omega_FL)+\frac{\alpha\eta u^r}{\mathcal{B}^r} (Eg_{t\phi} +Lg_{tt}) }
{N_{\rm co}\Delta\sin^2\theta -\frac{\alpha\eta u^r}{\mathcal{B}^r} \left(g_{t\phi}^2-g_{\phi\phi} g_{tt}\right)},
\end{align}
where $N_{\rm co}$ is the normalization of the co-moving frame defined in Eq.~\ref{eq:ncofac}. Finally, we can re-write $u^\theta$ in favor of $u^r$ with the relation $u^\theta=\frac{\mathcal{B}^\theta u^r}{\mathcal{B}^r}$ (which follows from conservation of $\eta$). 

Given conserved quantities $E,L,\eta,\Omega_F$, the four-velocity is thus completely determined by $u^r$, which can be computed from the normalization condition $u^\mu u_\mu=-1$. In fact, this normalization condition can be reduced to a quartic polynomial\footnote{The quartic can equivalently be cast in terms of $u_p\equiv \sqrt{u^ru_r+u^\theta u_\theta}$ (e.g. \citealp{takahashi_magnetohydrodynamic_1990}).} in $u^r$:\begin{align}
\label{eq:windeq}
    \tilde{A} (u^r)^4+\tilde{B}(u^r)^3+\tilde{C}(u^r)^2+\tilde{D}(u^r)+\tilde{E}=0.
\end{align}
The coefficients are\begin{align}
\tilde{A}&=\eta^2\kappa^2\lambda
\\\nonumber
\tilde{B}&=-2\alpha^{-1}\mathcal{B}^r\eta N_{\rm co}\kappa\lambda\Delta\sin^2\theta
\\\nonumber
\tilde{C}&=(\alpha^{-1}\mathcal{B}^r N_{\rm co}\Delta\sin^2\theta)^2\lambda+\eta^2\kappa\left[\kappa-(g_{tt}L^2+2g_{t\phi}EL+g_{\phi\phi}E^2)\right]
\\\nonumber
    \tilde{D}&=-2\alpha^{-1}\mathcal{B}^r\eta\kappa \Delta\sin^2\theta[N_{\rm co}-(E-\Omega_F L)^2]
    \\\nonumber
    \tilde{E}&=(\alpha^{-1}\mathcal{B}^r)^2N_{\rm co}^2\Delta\sin^2\theta[N_{\rm co}-(E-\Omega_F L)^2],
\end{align}
where we defined\begin{align}
    \kappa\equiv g_{t\phi}^2-g_{tt}g_{\phi\phi},\qquad \lambda\equiv g_{rr}+\left(\frac{\mathcal{B}^\theta}{\mathcal{B}^r}\right)^2g_{\theta\theta}.
\end{align}
Taking $u^r=0$ in Eq.~\ref{eq:windeq} gives the expression for $L$ employed in Eq.~\ref{eq:constraint}.

Throughout this paper, we compute the roots of the wind equation numerically. In practice, we find that two roots are always complex, while two are real, and the two real solution branches converge at the Alfv\`en point.

\subsection{Stagnation Surface}
\label{sec:stag}
In this section, we demonstrate the existence of the stagnation surface mentioned in \S\ref{sec:winds}. The stagnation surface represents the boundary between inflows and outflows, and its location in a Kerr background was first derived by \citet{takahashi_magnetohydrodynamic_1990}. This original derivation centers on regularity of the flow at the launch point. Here, we go one step further and derive the location of the stagnation surface directly from the equations of motion, without explicitly demanding regularity. 

To begin, we turn to conservation of $T^{\mu\nu}_{\rm MHD}$:\begin{align}
\label{eq:euler1}
    \nabla_\mu T^{\mu\nu}_{\rm MHD}&=0\Longrightarrow \nabla_\mu T^{\mu\nu}_{\rm Fluid}=-\nabla_\mu T^{\mu\nu}_{\rm EM}.
\end{align}
The term $\nabla_\mu T^{\mu\nu}_{\rm EM}$ is the Lorentz force density, which must be perpendicular to the magnetic field in the fluid frame. So letting $b^\mu\equiv u_\nu(\star F)^{\nu\mu}$ denote the fluid-frame magnetic field (following the notation of \citealp{gammie_harm_2003}), this perpendicularity constraint can be expressed covariantly as\begin{align}
    b_\nu\nabla_\mu T^{\mu\nu}_{\rm EM}=0,
\end{align}
which is also derived explicitly in Appendix D of \cite{chael2024hybrid}.
This implies, in the cold limit, that\begin{align}
\label{eq:euler2}
    b_\nu\nabla_\mu T^{\mu\nu}_{\rm Fluid}=\rho b_\nu u^\mu\nabla_\mu u^\nu=0.
\end{align}

Let us expand the above equation at the launch point. There, the fluid frame coincides with the co-rotating frame, which travels with four-velocity (Eqs.~\ref{eq:launcheq}-\ref{eq:ncofac})\begin{align}
    u_0^\mu=\frac{1}{\sqrt{N_{\rm co}}}(1,0,0,\Omega_F),\qquad N_{\rm co}\equiv -(g_{tt}+2g_{t\phi}\Omega_F+g_{\phi\phi}\Omega_F^2).
\end{align}

We will evaluate Eq.~\ref{eq:euler2} by re-writing the four-velocity $u^\mu$ in terms of its components in the co-rotating frame $u^{(a)}$. To perform the appropriate coordinate transformation, we use the tetrad construction briefly described in Section~\ref{sec:polarization}:\begin{align}
 u^\mu= e^\mu_{(a)}u^{(a)},
\end{align}
where the co-rotating tetrad is now explicitly given by\footnote{In the above expression, the Greek index refers to row and the latin index refers to column.} \citep{dexter_public_2016}\begin{align}
    e^\mu_{(a)}&=\begin{pmatrix}
        \frac{1}{\sqrt{N_{\rm co}}}&0&0&\frac{\Omega_F}{\sqrt{N_{\rm co}}}
        \\
        0&\frac{1}{\sqrt{g_{rr}}}&0&0
        \\
        0&0&\frac{1}{\sqrt{g_{\theta\theta}}}&0
        \\
        \frac{\Omega_Fg_{\phi \phi}}{\sqrt{N_{\rm co}\Delta \sin^2\theta}}&0&0&-\frac{g_{tt}}{\sqrt{N_{\rm co}\Delta\sin^2\theta}}
    \end{pmatrix}.
\end{align}
Thus, Eq.~\ref{eq:euler2} can be re-written as\begin{align}
    0&= b_\nu u^{(a)}e^{\mu}_{(a)}\nabla_{\mu}e_{(b)}^\nu u^{(b)}=b_\nu u^{(a)}e^{\mu}_{(a)}\left[\p_\mu (e_{(b)}^\nu u^{(b)})+\Gamma^{\nu}_{\mu\sigma}e^{\sigma}_{(b)}u^{(b)}\right],
\end{align}
where $\Gamma$ are the Christoffel symbols. 

At the launch point, we have $u=\p_{(t)}$ in the co-rotating frame by construction. However, poloidal derivatives of the four-velocity might diverge at the launch point. So for our purposes, we can throw out terms containing a spatial component of $u^{(a)}$ so long as they don't multiply a poloidal derivative. This means that\begin{align}
    0&=b_r[e_{(r)}^ru^{(r)}\p_r  (e^{r}_{(r)}u^{(r)})+e_{(\theta)}^\theta u^{(\theta)}\p_\theta  (e^{r}_{(r)}u^{(r)})+e_{(t)}^t e^{t}_{(t)} \Gamma^{r}_{tt}+e_{(t)}^\phi e^{\phi}_{(t)} \Gamma^{r}_{\phi\phi}+2e_{(t)}^t e^{\phi}_{(t)} \Gamma^{r}_{t\phi}]+r\leftrightarrow \theta.
\end{align}
Furthermore, the condition $u^r=u^\theta=0$ at the launch point implies that $\frac{b^r}{\mathcal{B}^r}=\frac{b^\theta}{\mathcal{B}^\theta}$ (Eqs. 16-17 of \citealp{gammie_harm_2003}), so we can replace the four-vector $b$ with the more familiar $\mathcal{B}$ in the above expression:\begin{align}
\label{eq:randomapp}
     0&=\mathcal{B}_r[e_{(r)}^ru^{(r)}\p_r  (e^{r}_{(r)}u^{(r)})+e_{(\theta)}^\theta u^{(\theta)}\p_\theta  (e^{r}_{(r)}u^{(r)})+e_{(t)}^t e^{t}_{(t)} \Gamma^{r}_{tt}+e_{(t)}^\phi e^{\phi}_{(t)} \Gamma^{r}_{\phi\phi}+2e_{(t)}^t e^{\phi}_{(t)} \Gamma^{r}_{t\phi}]+r\leftrightarrow \theta.
\end{align} 
Employing the MHD result that $u^\theta=\frac{\mathcal{B}^\theta}{\mathcal{B}^r}u^r$ and plugging into Eq.~\ref{eq:randomapp}, we obtain\begin{align}
    0&=\mathcal{B}_r\left\{e_{(r)}^r e^{r}_{(r)}[u^{(r)}\p_r  u^{(r)}+u^{(\theta)}\p_r u^{(\theta)}]+e_{(t)}^t e^{t}_{(t)} \Gamma^{r}_{tt}+e_{(t)}^\phi e^{\phi}_{(t)} \Gamma^{r}_{\phi\phi}+2e_{(t)}^t e^{\phi}_{(t)} \Gamma^{r}_{t\phi}\right\}+r\leftrightarrow \theta,
\end{align}
where $r\leftrightarrow \theta$ indicates that each $r$ index is swapped with a $\theta$ index.
Expanding out the Christoffel symbols gives
\begin{align}
    0&=
    \mathcal{B}_r\left\{e_{(r)}^r e^{r}_{(r)}[u^{(r)}\p_r  u^{(r)}+u^{(\theta)}\p_r u^{(\theta)}]-\frac{1}{2}g^{rr}\left(e_{(t)}^t e^{t}_{(t)} \p_r g_{tt}+e_{(t)}^\phi e^{\phi}_{(t)} \p_r g_{\phi\phi}+2e_{(t)}^t e^{\phi}_{(t)}\p_r g_{t\phi} \right)\right\}+r\leftrightarrow \theta
    \\&=
    \frac{1}{2}\mathcal{B}_rg^{rr}\left\{\frac{1}{2}\p_r[(u^{(r)})^2+(u^{(\theta)})^2]-N_{\rm co}^{-1}\left(\p_r g_{tt}+2\Omega_F\p_r g_{t\phi}+\Omega_F^2\p_rg_{\phi\phi}\right)\right\}+r\leftrightarrow \theta
    \\&=
    \label{eq:stageq2}
    \frac{1}{2}[(u^{(p)})^2+\log N_{\rm co}]',
\end{align}
where $(u^{(p)})^2\equiv (u^{(r)})^2+(u^{(\theta)})^2$ is the magnitude of the poloidal velocity in the co-rotating frame, and prime denotes differentiation along a poloidal fieldline (i.e. $f'\equiv \mathcal{B}^i\p_i f$ for an arbitrary function $f$). 

\begin{figure}[h]
    \centering
    \includegraphics[width=.48\textwidth]{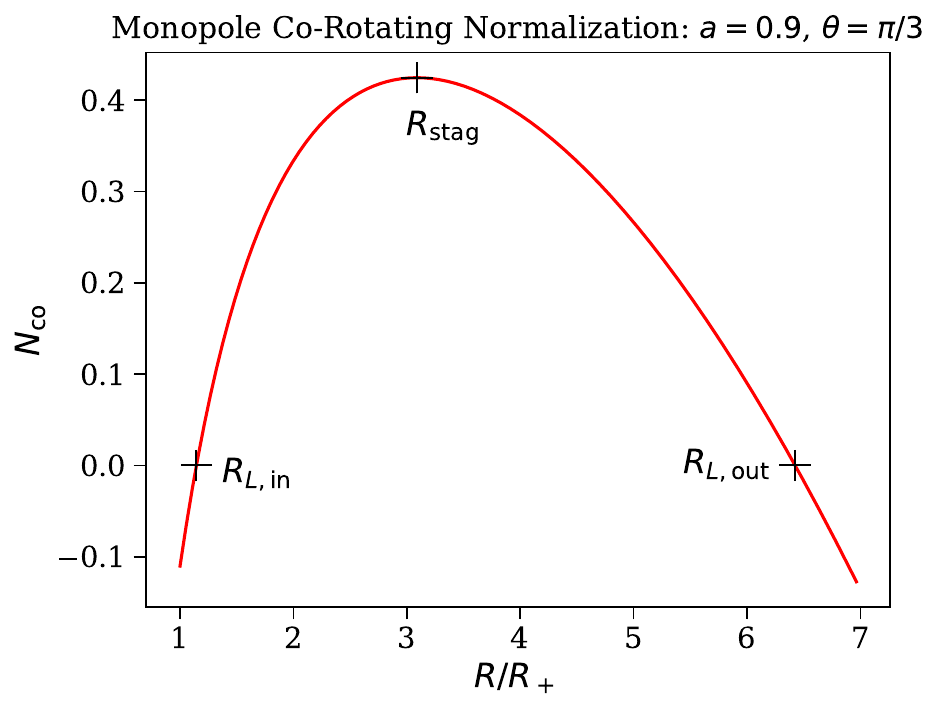}
    \caption{Normalization factor of four-velocity in co-rotating frame, plotted for $a=0.9$ and $\Omega_F=a/8$ along the monopolar fieldline $\theta=\pi/3$. The three crosses respectively refer to the inner light cylinder, the stagnation surface, and the outer light cylinder, and the radial coordinate is plotted in units of the event horizon outer radius $R_+=r_+\sin\theta$. The inner light cylinder appears only when the black hole spin is somewhat large $(a\gtrsim 0.5)$.}
    \label{fig:stag}
\end{figure}

For an outflow to be launched, $(u^{(p)})^2$ needs to grow as one moves radially outward along fieldline. Therefore, Eq.~\ref{eq:stageq2} shows that an outflow can be launched only when $N$ is decreasing along the fieldline, and an inflow can be launched when $N$ is increasing. The stagnation surface occurs at the boundary, where $ N'=0$. We plot both the light cylinders and the stagnation surface for a sample monopolar fieldline in Figure~\ref{fig:stag}. 

\begin{figure}[h]
    \centering
    \includegraphics[width=.48\textwidth]{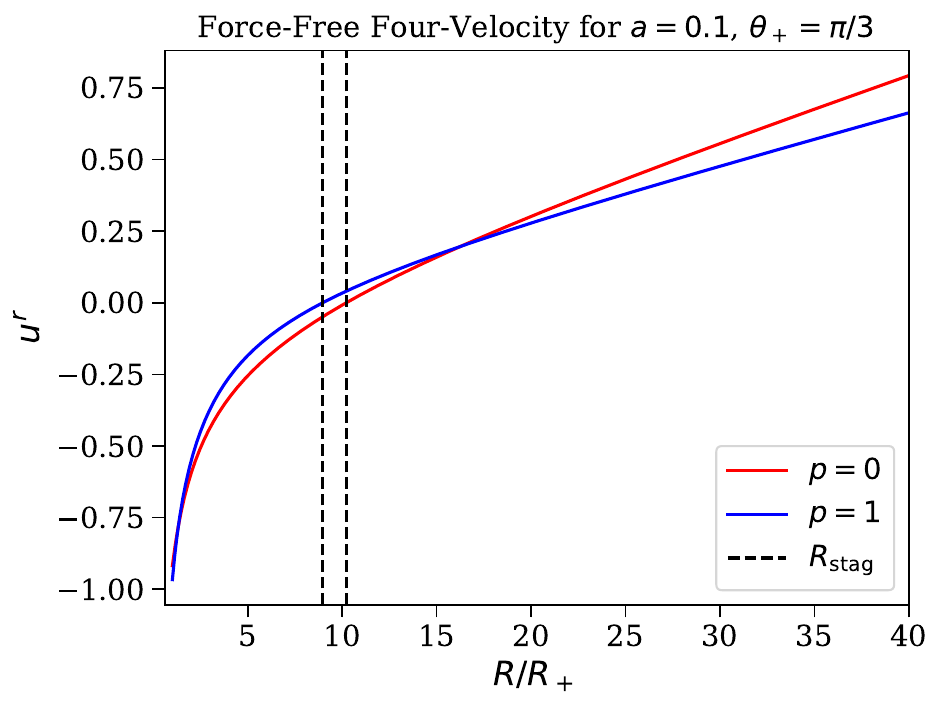}
    \caption{Force-free winds with $a=0.1$ that are launched (i.e. satisfy $u^r=0$) at their respective stagnation surfaces, shown in dashed lines. Both the monopole and paraboloid fieldlines are chosen so that they intersect the horizon off the equatorial plane at $\theta=\pi/3$.}
    \label{fig:inflow}
\end{figure}

By placing the launch point of a wind precisely at the stagnation surface, one can ensure a smooth transition from inflow to outflow. Indeed, in Figure~\ref{fig:inflow}, we plot the flows of two sample winds (monopole and paraboloid) all the way down to the horizon. We can see that the flows are smooth both at the horizon and at the stagnation surface, where the sign of the radial four-velocity flips.

\subsection{Critical Conditions}
\label{sec:critapp0}
In this section, we derive the critical conditions presented in \S\ref{sec:critsec}.
The need for explicit regularity conditions at the critical points arises because the expression for the poloidal acceleration contains apparent poles there \citep{takahashi_magnetohydrodynamic_1990}:\begin{align}
\label{eq:accel}
    \frac{d\log u_p}{dR}&= \frac{N}{D},\quad D\propto (u_p^2-u_a^2)^2(u_p^2-u_F^2)(u_p^2-u_S^2),
\end{align}
where $u_p\equiv\sqrt{u_ru^r+u_\theta u^\theta}$ is the poloidal four-velocity, $u_a$ is the Alfv\'en speed, $u_F$ is the fast magnetosonic speed, and $u_S$ is the slow magnetosonic speed. In cold MHD, the slow speed becomes zero, so it can only produce a discontinuity at the launch point where $u_p=0$. But at the other two critical points ($u_p=u_a$ and $u_p=u_F$), one must demand a vanishing residue ($N=0$) so the acceleration does not diverge. Fortunately, it turns out regularity at the fast point implies regularity at the Alfv\'en point, so we only need to concern ourselves with the former \citep{li_thesis}.  

At the fast point, one must be cautious, as several classes of trans-fast outflows (winds that smoothly cross the fast magnetosonic point) exist. For the monopole $(p=0)$, this is a well-studied problem, and \cite{takahashi_cold_1991} provides many examples of monopolar winds with fast points \emph{inside} the light cylidner (i.e. $R_F\Omega_F<1$); these winds are launched with high rotational velocities that provide most of the energy for poloidal acceleration. However, we are interested in winds that are mainly sourced by electromagnetic (Poynting) energy, as this type of flow is what will ultimately match onto the Blandford-Znajek model. This direction of energy conversion can be achieved when the fast point is located \emph{outside} the light cylinder. When the fast point is located at spatial infinity, the critical conditions depend on whether the collimation parameter $p$ is equal to zero or greater than zero. We discuss the two cases below.

\subsubsection{Fast Point at Infinity With $p=0$}
\label{sec:critapp}
Because we put the critical point at spatial infinity, then we can work in flat space to derive the critical conditions. In flat space, the lapse is $\alpha=1$ and the wind equation becomes
 \citep{takahashi_cold_1991}:\begin{align}
\label{eq:wind}
    &1+u_p^2=E^2\frac{(1-R^2)(1-L\Omega_F/E)^2-2(1-L\Omega_F/E)^2M^2_p-(L^2\Omega_F^2/E^2-R^2)M_p^4/R^2}{(1-R^2-M^2_p)^2},
\end{align}
where $R\equiv r\Omega_F\sin\theta$ is the dimensionless cylindrical radius, and $M_p^2\equiv u_p^2/u_a^2$ is the poloidal Alfv\'en Mach number with $u_a=\mathcal{B}_p^2/\rho$ the poloidal Alfv\'en four-velocity. Using conservation of mass loading $\eta$, we can rewrite the Mach number as\begin{align}
\label{eq:mpeq}
    M_p^2&=\frac{u_p^2}{\mathcal{B}_p^2/\rho}=\frac{u_p}{\mathcal{B}_p/\eta}=\frac{u_pR^2}{\sigma_M\sin^2\theta}.
\end{align}
At the fast point, expanding $D=0$ in large $R$ and plugging in Eq.~\ref{eq:mpeq} for the Mach number (see Eqs.~12-14 of \citealp{takahashi_trans-fast_1998} for the full expressions of $N$ and $D$) gives:\begin{align}
\label{eq:expand}
    u_F&=E^{2/3}(\sigma_M\sin^2\theta)^{1/3}-\sigma_M\sin^2\theta+\mathcal{O}(R^{-2}),
\end{align}
where $u_F$ is the poloidal four velocity at the fast point. Similarly, expanding the wind equation at large $R$ tells us that\begin{align}
\label{eq:expand2}
    E^2=\frac{(\sigma_M\sin^2\theta+u_F)^2(1+u_F^2)}{u_F^2}+\mathcal{O}(R^{-2}).
\end{align}
Solving~Eqs.~\ref{eq:expand} and \ref{eq:expand2} as $R\to \infty$ gives the critical relations at infinity:\begin{align}
\label{eq:criteq0}
    E^2=(1+u_F^2)^3,\qquad u_F=(\sigma_M\sin^2\theta)^{1/3},
\end{align}
in agreement with \cite{michel_relativistic_1969} and \cite{goldreich_stellar_1970}. Noting that at as $r\to\infty$, the azimuthal velocity approaches zero (else the angular momentum of the fluid would diverge), so $\gamma_\infty=\frac{1}{\sqrt{1-v_p^2}}$ and hence\begin{align}
 E=\gamma_\infty^3,\quad   \sigma_M&=\frac{(\gamma_\infty^2-1)^{3/2}}{\sin^2\theta},
\end{align}
where $\gamma_\infty$ is the terminal Lorentz factor.

\subsubsection{Fast Point at Infinity With $p>0$}
\label{sec:critapp2}
In addition to the dimensionless cylindrical radius $R\equiv r\sin\theta\Omega_F$ used in the $p=0$ case, we will also use a dimensionless vertical coordinate $Z\equiv r\Omega_F\cos\theta$ here. In terms of these coordinates, the magnetic fields are:\begin{align}
    A_\phi&\equiv \eta\sigma_M \Omega_F^{-2} (R^2+Z^2)^{p/2}\left(1-\frac{Z}{\sqrt{Z^2+R^2}}\right),\quad
    B^R=-\frac{A_{\phi,Z}}{\sqrt{-g}},\quad B^Z=\frac{A_{\phi,R}}{\sqrt{-g}}
\end{align}
Since the fast point is at infinity, we can again evaluate these expressions explicitly in flat space, where $\sqrt{-g}=R$ and hence\begin{align}
    B^R&=\frac{\eta\sigma_M\Omega_F^{-2}}{R}(R^2+Z^2)^{\frac{p-3}{2}}\left[R^2+pZ\left(Z-\sqrt{R^2+Z^2}\right)\right]
    \\
    B^Z&=\eta\sigma_M\Omega_F^{-2}(R^2+Z^2)^{\frac{p-3}{2}}(Z-pZ+p\sqrt{R^2+Z^2}).
\end{align}
So we see that $R$ can be expressed as an analytic function of $Z$ as $r\to\infty$:\begin{align}
    Z=r\Omega_F\cos\theta\to r\Omega_F,\qquad R=r\sin\theta\Omega_F\to \sqrt{\frac{2\psi r^{2-p}\Omega_F^{2}}{\eta\sigma_M }}=\sqrt{\frac{2\psi \Omega_F^{2} Z^{2-p}}{\eta\sigma_M}},
\end{align}
which gives the asymptotic scaling $Z\propto R^{\frac{2}{2-p}}$ presented in Eq.~\ref{eq:scaleeq}.
Since $0<p<2$, then the large $r$ limit is equivalent to $Z\gg R$, giving the following magnetic fields\begin{align}
    B^R\to \frac{1}{2}(p-2)\Omega_F^{-2}\eta\sigma_M Z^{p-3}R\to \frac{1}{2}(p-2)\sqrt{2\psi\eta\sigma_M}\Omega_F^{-1}Z^{p/2-2},\qquad B^Z\to \eta \sigma_M\Omega_F^{-2} Z^{p-2},
\end{align}
from which we see that the magnetic flux function $\Phi$ approaches a constant:\begin{align}
\label{eq:magneticflux}
    \Phi\equiv R^2B_p\to R^2B_Z\to 2\psi.
\end{align}
This is important because the expression for $D$ generically contains a term of the form $(\log \Phi)'$ \citep{takahashi_trans-fast_1998}, which will vanish here.
Then defining $A\equiv \frac{2\psi\Omega_F^2}{\eta\sigma_M}$ and solving $D=0$ at large $Z$ gives\begin{align}
    D&=0=\left[\frac{AE^2u_p^2}{(A\sigma_M+u_p)^2}-\frac{u_p^2(A\sigma_M+u_p)}{\sigma_M}\right]Z^{2-p}+...
\end{align}
Then the wind equation in this limit is \begin{align}
    0&=E^2-\frac{(A\sigma_M+u_p)^2(1+u_p^2)}{u_p^2},
\end{align}
which are the same two conditions as the monopole except with $\sigma_M\sin^2\theta\to A\sigma_M$. Thus the modified critical conditions are:\begin{align}
\label{eq:criteq1}
    E=(1+u_F^2)^{3/2}=\gamma_\infty^3,\qquad u_F=(A\sigma_M)^{1/3}=\left(\frac{2\psi\Omega_F^2}{\eta}\right)^{1/3}.
\end{align}
Then letting $\{r_{0},\theta_{0}\}$ denote the launch point of the fluid, we have\begin{align}
    \psi=\eta\sigma_M\Omega_F^{-2}  \left(\frac{R_{0}}{\sin\theta_0}\right)^p(1-\cos\theta_0),
\end{align}
from which we can compute $\sigma_M$: \begin{align}
    \sigma_M&=\frac{\sin^p\theta_0(\gamma_\infty^2-1)^{3/2}}{R_{0}^p(1-\cos\theta_{0})}.
\end{align}
Now, one might hope that unique values of $\gamma_\infty$ or $R_0$
could be isolated by imposing the additional constraint
that $N=0$ at the fast-point (our derivations have so
far exploited only $D=0$). However, $N$ always has the
leading order scaling $N\sim r^{-3}$, so imposing $N=0$ would
be redundant when the fast point is at infinity.

\subsubsection{Fast Point at Finite Radius}
\label{app:finitefms}
The choice to put the fast point at infinity for monopolar and paraboloidal fieldlines is robust, as the scaling $E\sim \sigma_M\sim\gamma_\infty^3$ is known to hold for monopolar jets even when the fast point moves to a finite radius \citep{goldreich_stellar_1970,beskin_mhd_1998}. This scaling relation represents inefficient energy conversion and is actually reflective of a much broader class of fieldline configurations. Indeed, the efficiency of energy conversion depends solely on the magnetic flux function $\Phi$ defined in Eq.~\ref{eq:magneticflux}; one can show that efficient acceleration (i.e. $E\sim\gamma_\infty)$ is possible if and only if $\Phi$ decreases along fieldlines \citep{begelman_asymptotic_1994,tchekhovskoy_efficiency_2009}. For the class of stream functions considered in this paper, $\Phi$ approaches a constant at large $r$, meaning efficient acceleration is impossible, and moving the fast point to a finite radius would not affect the relationship between $E$ and $\gamma_\infty$.  

However, the GS equation is known to admit other solutions that \emph{do} support a decreasing magnetic flux, thus enabling efficient energy conversion. Since the relation $E\sim\gamma^3$ will always hold precisely at the fast point \citep{beskin2006effective}, this efficient energy conversion will begin \emph{beyond} the fast point, particularly through processes like the ``magnetic nozzle" effect \citep{camenzind1989hydromagnetic,li_electromagnetically_1992,nakamura2013parabolic}. Dissipative effects like magnetic reconnection can cause acceleration beyond the fast point too \citep{coroniti_magnetically_1990}. Indeed, one can modify the stream function in Eq.~\ref{eq:streamfunc} to include a term of the form\begin{align}
    \psi\to\psi+\epsilon r\sin\theta,
\end{align}
which still represents an approximate solution to the GS equation if $\epsilon$ is small, and also allows for $\Phi$ to decrease when $\epsilon>0$ \citep{beskin2006effective,pu2015steady,ogihara_mechanism_2019}. 

In addition to a direct modification of the stream function, \cite{takahashi_constraints_2008} proposed another way to approximately model the effects of efficient energy conversion. In their model, one specifies $\overline{B}(r)$ as an \emph{input}, from which a smooth velocity profile emerges that passes through a finite fast point. This model was successfully employed in the work of \cite{pu_properties_2020} to characterize trans-fast MHD flows. However, this model is not particularly helpful for our work, as it is more difficult to make a clean connection to the force-free limit: if $\overline{B}$ is a free model parameter, then it is not obvious why it must become constant as $\sigma_M\to\infty$ (Eq.~\ref{eq:bbarconstant}). Furthermore, we are able to achieve the exact same plasma dynamics despite using a different set of critical conditions; our Figure~\ref{fig:testcap} is qualitatively similar to Figure 3 in \cite{pu_properties_2020}. Polarization measurements similarly should not differ substantially between our model and that of \cite{pu_properties_2020}, as their choice for $\overline{B}$ differs from ours only by a $\mathcal{O}(\gamma_\infty^{-2})$ correction far away from the black hole (compare our Eq.~\ref{eq:radcorrection} to their Eq.~15).

\subsection{MHD Correction to Radiation Condition}
\label{app:radiationcondition}
Here, we derive the MHD correction to the radiation that is presented in Eq.~\ref{eq:radcorrection}. To do so, we begin by combining Eqs.~\ref{eq:bbarsolve} and \ref{eq:usolve} in the flat space ($r\to\infty,\alpha\to 1$) limit, which allows us to express the azimuthal magnetic field as\begin{align}
   \lim_{r\to \infty} \overline{B}&=-\frac{ E\eta\Omega_Fr^2\sin^2\theta}{\Omega_F^2r^2\sin^2\theta+\eta u^r/\mathcal{B}^r}.
\end{align}
Expanding, we have in the large $r$ limit that\begin{align}
    \mathcal{B}^r&\to\eta\sigma_M\Omega^{p-2}r^{p-2},\qquad r^2\sin^2\theta\to \begin{cases}
        r^2(2\psi_0-\psi_0^2),&p=0
        \\
        2\psi_0 r^{2-p},&p>0
    \end{cases},
\end{align}
where $\psi_0=r^p(1-\cos\theta)$ is the fixed poloidal fieldline. Therefore, \begin{align}
    \lim_{r\to\infty}\overline{B}&=
    \begin{cases}
    -\frac{\psi_0 E\eta\Omega_F^{-1}(2-\psi_0) }{\psi_0(2-\psi_0)+\sigma_M^{-1}u^r},&p=0
    \\
    -\frac{2\psi_0 E\eta\Omega_F^{-1} }{2\psi_0+\Omega_F^{-p}\sigma_M^{-1}u^r},&p>0.
    \end{cases}
\end{align}
In the force-free limit, the second term in the denominator goes to zero as $\sigma_M\to \infty$. But in MHD, we must keep it. Assuming, in that case, that $u^r$ has asymptoted to its terminal value of $u^r\to\sqrt{\gamma_\infty^2-1}$, then we apply the critical conditions (Eqs.~\ref{eq:monocrit}-\ref{eq:parcrit}) to find that the two cases merge, and\begin{align}
    \lim_{r\to\infty}\overline{B}&=
        -\eta\Omega_F^{-1}\gamma_\infty\left(\gamma_\infty^2-1\right).
\end{align}
Meanwhile, the $\theta$ component of the electric field comes out to 
\begin{align}
     \lim_{r\to\infty} r^2\sin\theta \mathcal{E}^\theta=-\sin\theta F_{t\theta}=-r^2\sin^2\theta\Omega_F\mathcal{B}^r=-\eta\sigma_M\Omega_F^{p-1}r^{p}\sin^2\theta=-\eta\Omega_F^{-1}(\gamma_\infty^2-1)^{3/2},
\end{align}
where we expressed $\sigma_M$ in terms of $E$ using the critical conditions and once again found that the final result doesn't depend on $p$. With these two results, we find the MHD radiation condition\begin{align}
    \lim_{r\to\infty}\frac{r\sin\theta \mathcal{B}^\phi}{r \mathcal{E}^\theta}=\lim_{r\to\infty}\frac{\overline{B}}{r^2\sin\theta \mathcal{E}^\theta}=\frac{\gamma_\infty}{\sqrt{\gamma_\infty^2-1}}=1+\frac{1}{2\gamma_\infty^2}+\mathcal{O}(\gamma_\infty^{-4}).
\end{align}

\section{Ray Tracing}
\label{app:raytrace}
Here, we derive the observed specific intensity on the final image, combining the influences of radiative transfer, Doppler boosting, and gravitational redshifting. If the jet wall traces out the contour $\psi={\rm const}$, then the vector $\p_\mu\psi$ will be normal to the jet wall. Thus in the emitter frame, the vector normal to the jet wall, $w$, is\begin{align}
    w^{(a)}&=\eta^{(a)(b)}e^{\mu}_{(b)}\partial_\mu\psi.
\end{align}
Hence the path length $\ell_p$ will be\begin{align}
    l_p&=\frac{p^{(t)}}{p^{(i)}\hat{w}_{(i)}}W,\qquad \hat{w}^{(i)}\equiv \frac{w^{(i)}}{\sqrt{w^{(i)}w_{(i)}}},
\end{align}
where $W$ is the width of the jet wall in the emitting frame. We will assume $W$ scales linearly with the cylindrical emission radius $R_{\rm emit}$, meaning that the emitted specific intensity for spectral index $\alpha_\nu$ in the optically thin regime is\begin{align}
    I_{\nu,\rm emit}&=\frac{p^{(t)}}{p^{(i)}\hat{w}_{(i)}}\rho R_{\rm emit}\nu^{-\alpha_\nu}(|\vec{B}|\sin\theta_B)^{\alpha_\nu+1}.
\end{align}
From here, the \emph{observed} specific intensity $I_{\nu,\rm obs}$ needs an additional correction that comes from Doppler boosting and gravitational redshifting. If the source emits with spectral index $\alpha_\nu$, then conservation of $I_{\nu}/\nu^3$ implies that the observed intensity will be proportional to $g^{3+\alpha_\nu}I_{\nu,\rm emit}$, where the Doppler factor $g$ is \citep{narayan_polarized_2021}\begin{align}
    g&=\frac{1}{p^{(t)}}.
\end{align}
Combining all of these effects, the observed intensity will be proportional to \begin{align}
\label{eq:radgeneral}
    I_{\nu,\rm obs}&\propto \frac{g^{2+\alpha_\nu}}{p^{(i)}\hat{w}_{(i)}}\rho R_{\rm emit}\nu^{-\alpha_\nu}(|\vec{B}|\sin\theta_B)^{\alpha_\nu+1}.
\end{align}
Throughout the main text, we employ this result with $\alpha_\nu=1/2$, as appropriate for the M87 jet at optically thin frequencies.

We emphasize that Eq.~\ref{eq:radgeneral} holds specifically for a power-law EDF, where $\alpha_\nu=\frac{s-1}{2}$ is a constant. In contrast, a thermal EDF will produce more fluctuations in pitch angle dependence due to the exponential cutoff in the distribution function.

We can now approximate Eq.~\ref{eq:radgeneral} both inside and outside the light cylinder to derive how the intensity scales with impact parameter. To do so, we will assume that $r$ is sufficiently large so that we can adopt the relationship between $b$ and $r$ in Eq.~\ref{eq:impactparam}. Then inside the light cylinder, we take $g$ and $\ell_p$ to be constant. Outside the light cylinder, however, the plasma accelerates and we have asymptotically that\begin{align}
    g\sim r^{p/2}\propto b^{\frac{p}{2-p}},\qquad \frac{1}{p^{(i)}\hat{w}_{(i)}}\propto r^{p/2}\propto b^{\frac{p}{2-p}}.
\end{align}
Plugging these relations into Eq.~\ref{eq:radgeneral}, we find to leading order that the intensity scaling does not depend on $p$, and \begin{align}
\label{eq:intensity2}
    I(b)&\sim\begin{cases}
      b^{-5-2\alpha_\nu},R\ll R_L
      \\
      b^{-2-\alpha_\nu},R\gg R_L.
    \end{cases}
\end{align}
For our fiducial choice of $\alpha_\nu=0.5$, we therefore expect $I\sim b^{-6}$ and $I\sim b^{-2.5}$ inside and outside the light cylinder respectively. We note that the latter scaling ($b^{-2.5}$) is steeper than most intensity gradients measured in blazar observations \citep{burd2022dual}.



\section{Model Extensions}
Here, we discuss how changes to our model assumptions affect the predicted polarization pattern.
\subsection{Inclined Images}
\label{app:incline}
While all of our polarization results thus far have assumed a purely face-on observer inclination $(i=0^\circ)$, we will demonstrate that similar conclusions still hold when the observer is pushed to a small but nonzero angle off the jet axis. Since the M87 jet is believed to be viewed at $\sim 17^\circ$, we will adopt $i=17^\circ$ as our fiducial observer inclination in this Appendix \citep{mertens_kinematics_2016,walker_structure_2018,eht_paper5}.

\begin{figure}[h]
    \centering
    \includegraphics[width=.4\textwidth]{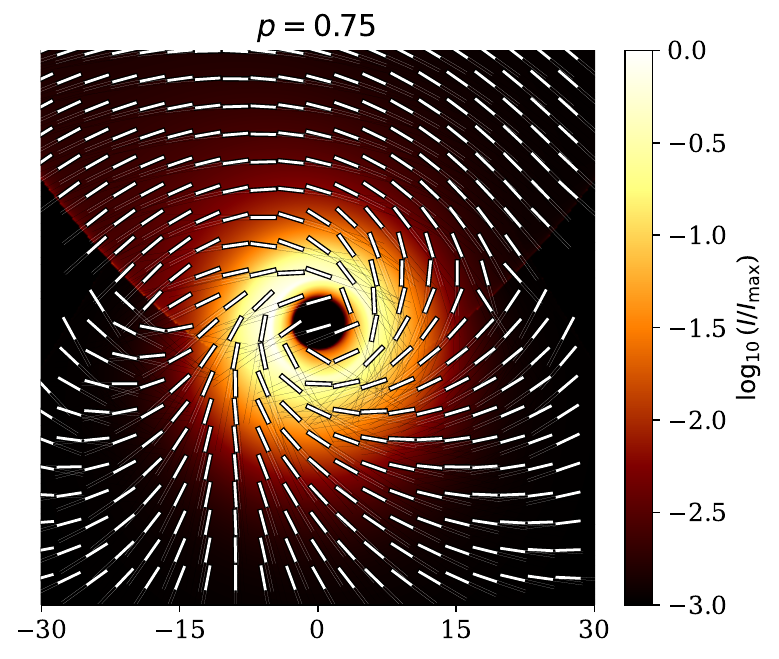}
    \includegraphics[width=.4\textwidth]{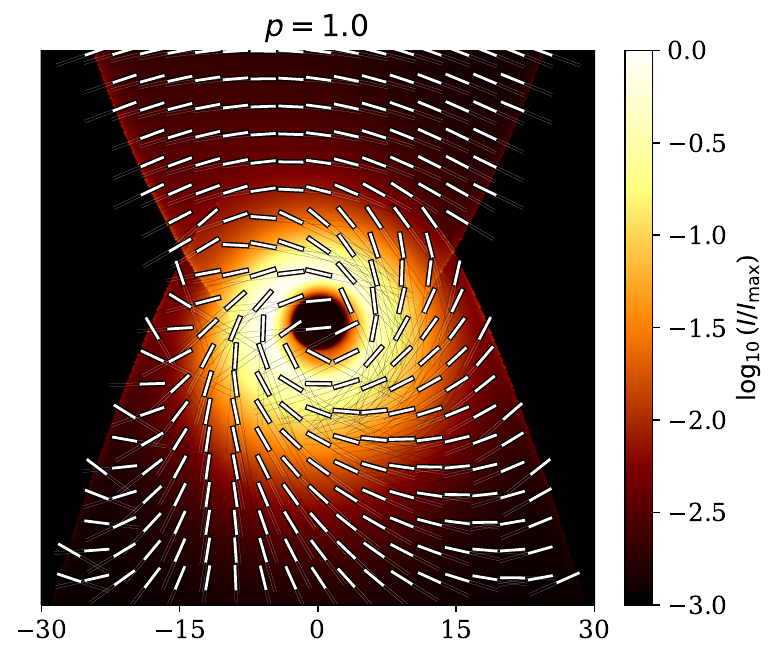}
    \caption{Images of $a=0.5$ force-free jets viewed at a $17^\circ$ inclination. The scale is in units of $M$, and the forward jet is in the upper half of the image with the counter-jet below.}
    \label{fig:incimage}
\end{figure}

We perform the same ray tracing procedure described in \S\ref{sec:imagesec} to generate the inclined images. Results are shown in Figure~\ref{fig:incimage} for two force-free $a=0.5$ jets. In these images, the forward (top) and counter (bottom) jets are no longer identical, and there are parts of the image that are completely dark since since they lie outside the lensed jet outline. Aberration now only affects the forward jet, so some of the polarization swings discussed in this paper will be weaker.

\begin{figure}[h]
    \centering
    \includegraphics[width=.9\textwidth]{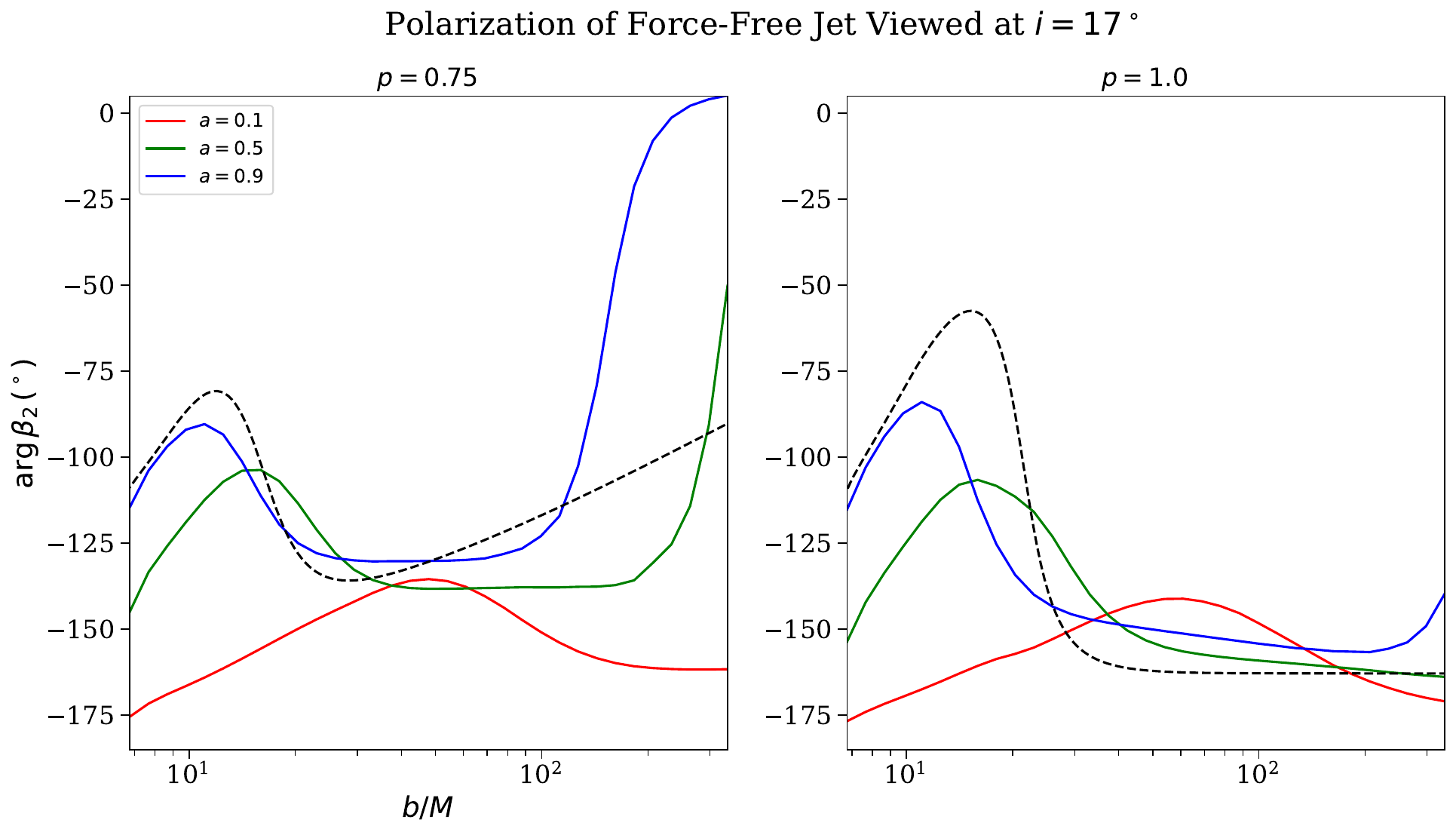}
    \caption{Polarization curves for $i=17^\circ$ force-free jets. The dashed curve, presented for comparison, is the polarization of the $a=0.9$ face-on jet.}
    \label{fig:incimage2}
\end{figure}

However, the signature still persists. We can compute ${\rm arg}(\beta_2)$ explicitly by integrating over circles on the inclined image (Eq.~\ref{eq:betadef}), which gives us the polarization curves in Figure~\ref{fig:incimage2}.
We see that in both the $p=0.75$ and $p=1.0$ jets, $\arg(\beta_2)$ displays the same characteristic spin-dependent swings. Moreover, the sub-paraboloidal fieldlines swing towards a radial polarization far from the hole, consistent with the interpretation presented in \S\ref{sec:collimation}. Curiously, the $p=1$ high-spin fieldline swings towards radial polarization too. This may be a finite inclination effect, wherein aberration stops far from the jet since the material is no longer beamed directly towards the observer.

There may be better metrics than $\arg(\beta_2)$ to capture the polarization structure in inclined images far from the black hole, where the image structure is no longer approximately axisymmetric. Indeed, we cannot delineate a single light cylinder on these one-dimensional plots of $\arg(\beta_2)$, as the light cylinder no longer corresponds to a specific impact parameter. However, it is clear that the spin dependence of the polarization swing still offers ample opportunities for spin measurements, and we will explore specific strategies to measure spin from these inclined observations in future work. 

\begin{figure}[h]
    \centering
    \includegraphics[width=.4\textwidth]{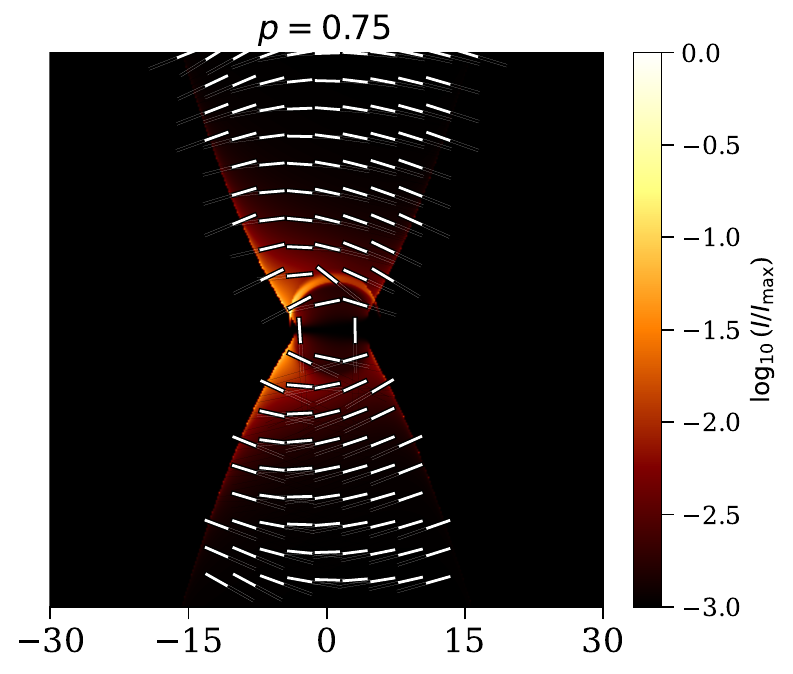}
    \includegraphics[width=.4\textwidth]{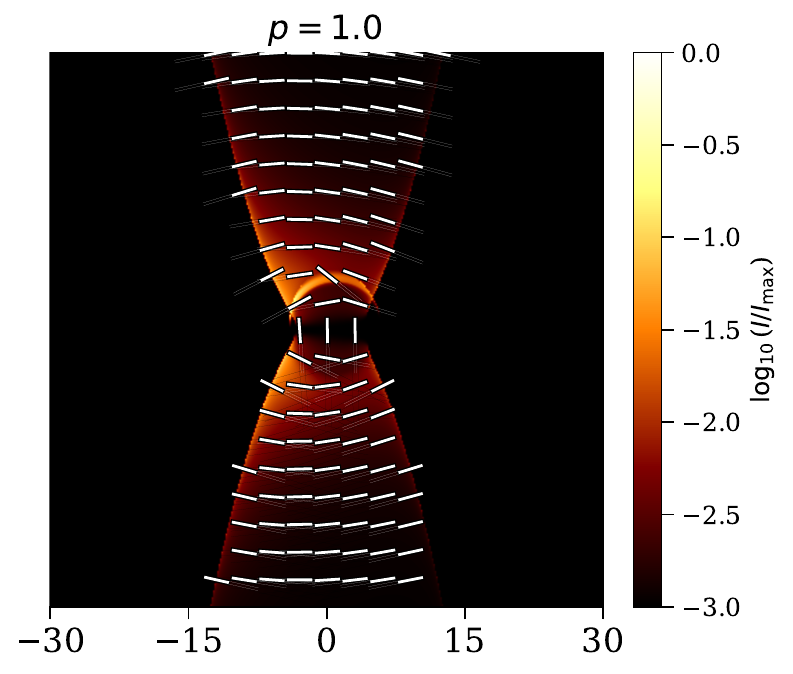}
    \caption{Images of $a=0.5$ force-free jets viewed at an inclination of $i=89^\circ$.}
    \label{fig:highincfig}
\end{figure}

In addition to the low-inclination regime, we expect our method to generalize to the high-inclination regime too. In Figure~\ref{fig:highincfig}, we show images of the same two jets ($p=0.75$ and $p=1.0$), but with the observer placed at $i=89^\circ$. 

\subsection{Off-Equatorial Fieldlines}
\label{sec:offeqapp}
Up until now, we have ray traced along the fieldlines that intersect the horizon in the equatorial plane. That is, we take $\psi_{\rm jet}=r_+$ (Eq.~\ref{eq:psijeteq}). In this section, we will explore the effects of varying $\psi_{\rm jet}$.

\begin{figure}[h]
    \centering
    \includegraphics[width=\textwidth]{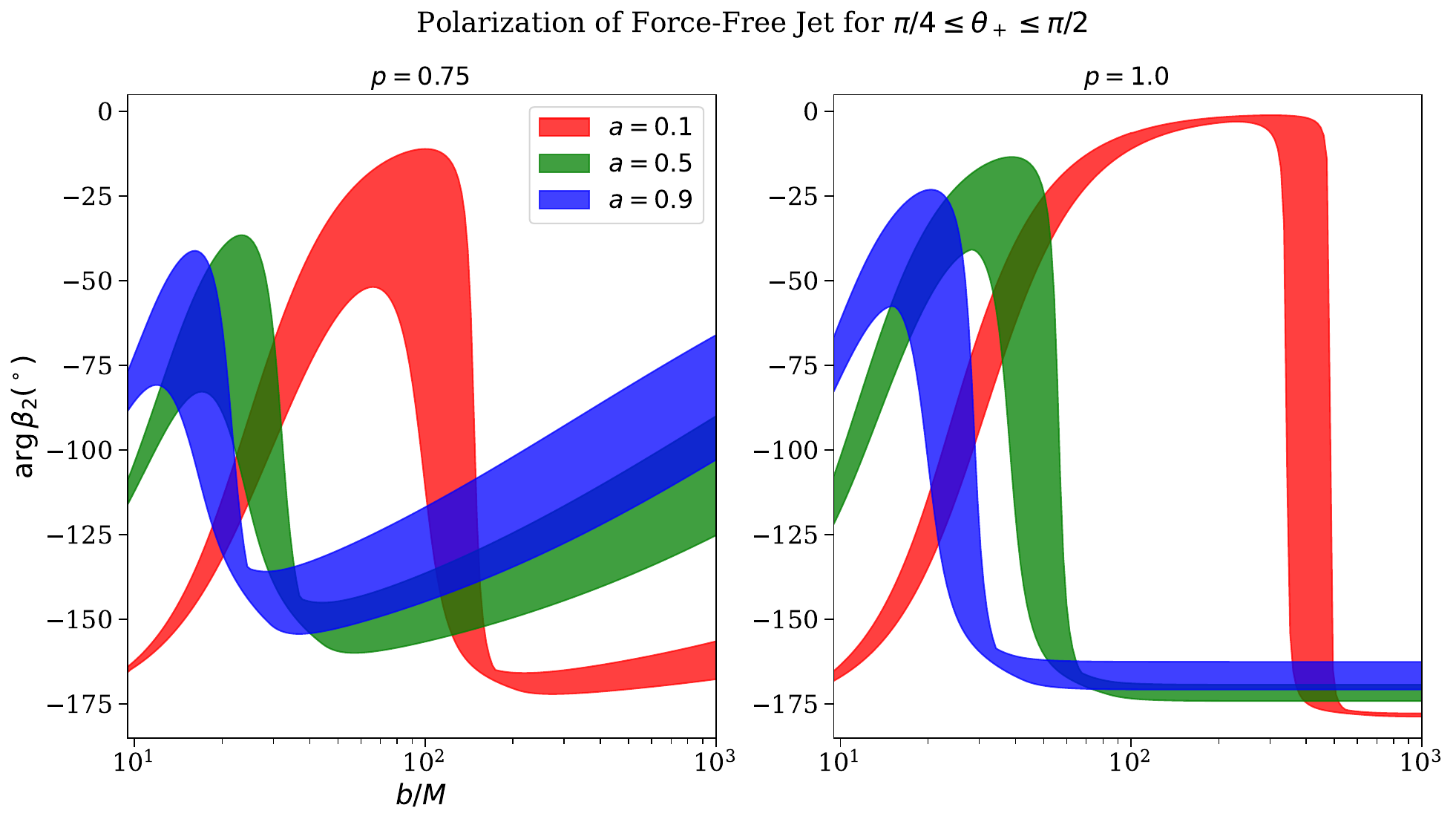}
    \caption{Ribbon plots show the range of observed polarization for force-free fieldlines that thread the horizon between $\theta_+=\pi/4$ and $\theta_+=\pi/2$. The thickness of the ribbon comes from differences in $\Omega_F$, as well as changes in aberration due to different jet opening angles.}
    \label{fig:polribbon}
\end{figure}

We plot in Figure~\ref{fig:polribbon} the polarization curves for a range of values of $\psi_{\rm jet}$ in our force-free model. In particular, we ray trace fieldlines that intersect the horizon at polar angles ranging from $\theta_+=\pi/4$ to $\theta_+=\pi/2$. As the fieldlines are pushed off the equatorial plane, the effects of aberration increase since the jet becomes more collimated and beamed towards the observer. According to Eq.~\ref{eq:omega2}, decreasing $\theta_+$ also increases $\Omega_F$ through the Znajek condition, thus shrinking the light cylinder (by a factor of at most 1.5). These two effects combine to create a spread in the polarization ribbons plotted in Figure~\ref{fig:polribbon}.

Importantly, the shape and character of the polarization swings stays the same while varying $\psi_{\rm jet}$, and the location of the swings remains relatively constant. Our proposed method for spin measurement therefore still holds weight. Future observational studies will also be able to better constrain the jet opening angle (equivalent to $\psi_{\rm jet}$), thus enabling even more accurate modelling to predict spin.

\software{
\texttt{kgeo} \citep[][]{chael_kgeo_2023},
eht-imaging library \citep{chael_eht-imaging_2018}
, 
Numpy \citep{harris_array_2020},
Matplotlib \citep{hunter_matplotlib_2007}
}

\bibliography{references.bib,bib.bib}

\end{document}